\documentclass[aps,prx,10pt,twocolumn,superscriptaddress,nofootinbib,floatfix]{revtex4-2}

\usepackage{amsmath,amssymb,bm,mathtools,mathrsfs}
\IfFileExists{physics.sty}{\usepackage{physics}}{%
  \usepackage{braket}
  
}
\usepackage{microtype}
\usepackage{xcolor}
\usepackage[colorlinks=true, linkcolor=blue, citecolor=blue, urlcolor=blue]{hyperref}
\usepackage{graphicx,overpic}
\graphicspath{{assets/}}
\newif\ifreviewmarks
\reviewmarkstrue

\newif\ifrevisionmarks
\revisionmarksfalse
\definecolor{revisionred}{RGB}{170,20,30}
\DeclareRobustCommand{\rev}[1]{\ifrevisionmarks\textcolor{revisionred}{#1}\else#1\fi}
\newenvironment{revision}{\par\begingroup\ifrevisionmarks\color{revisionred}\fi}{\par\endgroup}
\begin{document}

\title{Discovering Hidden Predictive States in Kinetic Transport}

\author{Pengfei Zhu}
\affiliation{Bundesanstalt für Materialforschung und -prüfung (BAM), 12205 Berlin, Germany}

\date{\today}

\begin{abstract}
How do autonomous macroscopic transport laws emerge from kinetic dynamics, and which hidden degrees of freedom must survive coarse-graining? We address these questions by formulating kinetic state reduction as a problem of predictive sufficiency rather than conventional moment truncation. Starting from Boltzmann transport dynamics, we establish a quantitative connection between unresolved microscopic information, irreducible prediction error, and finite-horizon observability. Microscopic preparations with identical macroscopic observables but distinct subsequent evolution provide direct witnesses of incomplete dynamical closure. We develop an entropy-constrained learned state-reduction framework, demonstrated here through linear projections, that identifies the additional predictive states without prescribing their physical identities and determines their necessity through independent response tests, state deletion, and dynamical consistency. An exactly solvable three-velocity model demonstrates that a hidden higher-order moment can influence future energy transport despite identical initial energy and heat flux. Numerical experiments recover this missing direction and reveal how the required state dimension depends on observation scale, prediction horizon, preparation ensemble, and accuracy. In a synthetic 32-channel Boltzmann system, the learned representation identifies predictive combinations spanning multiple kinetic branches and relaxation groups that are not captured by low-order global moments. We further distinguish observable prediction accuracy from autonomous state closure and quantify the information retained through unresolved memory. These results establish an operational framework for testing the emergence of effective transport laws and identifying the microscopic information required for their predictive validity.
\end{abstract}

\maketitle

\section{Introduction}

A central problem of statistical mechanics is to explain how autonomous macroscopic laws emerge from microscopic dynamics containing vastly more degrees of freedom~\cite{Lebowitz1999,Barnett2023,Lebowitz1988}. The Boltzmann equation describes the evolution of a distribution over phase space, whereas hydrodynamic and diffusive models propagate only a small set of fields~\cite{Bertin2006,Ginzburg2007,Thuroff2014}. Classical kinetic reductions, including the Chapman--Enskog expansion and Grad's moment method, establish systematic connections between these descriptions when the relevant asymptotic assumptions apply~\cite{ChapmanCowling1970,Grad1949}. Yet the selection of macroscopic variables remains a dynamical question: which microscopic distinctions may be discarded without compromising future predictions? In this work, we address that question operationally, for prescribed observables, preparation ensembles, spatial resolutions, and prediction horizons.

Heat transport illustrates why this distinction matters~\cite{Zhu2026JAP,Zhu2026NonHermitian,Zhu2026Snell}. Fourier's law yields an effective diffusion equation in an appropriate hydrodynamic regime, but kinetic transport can exhibit quasiballistic propagation, relaxation of heat flux, and multiple scattering times~\cite{Chen2021,Guyer1966}. Extended descriptions, including Cattaneo--Vernotte-type flux relaxation and phonon hydrodynamics, retain additional dynamical structure beyond the temperature field~\cite{Cattaneo1958,Guyer1966}. Their applicability depends on the collision operator, observation times, length scales, and initial preparations; they do not form a universal, exact hierarchy of successive kinetic limits~\cite{Chen2021,Guyer1966}. Rather than choosing in advance between Fourier, flux-relaxation, and higher-moment closures, one may ask how many independent state variables are actually necessary to reproduce specified transport observables to a given accuracy.

This question belongs to the broader theory of coarse-graining and reduced dynamics~\cite{Kabernik2018,Rahav2007,Feret2009}. Statistical projection methods developed by Mori and Zwanzig show that eliminating unresolved variables generally produces a memory kernel and a contribution from unresolved initial conditions~\cite{Mori1965,Zwanzig1961}. Thus, a low-dimensional observable is not necessarily a closed Markovian state, even when its microscopic evolution is deterministic. Systematic coarse-grained modeling likewise emphasizes that the adequacy of an effective description depends on the quantities and distributions it is designed to reproduce~\cite{Noid2013}. For transport, the distinction between discarding a rapidly relaxing variable and retaining its influence through memory is especially important: a reduced model may store the same predictive information either in additional instantaneous coordinates or in a suitably initialized history-dependent representation~\cite{Mori1965,Zwanzig1961,Rupe2022}. These choices must be compared at a matched information and storage budget.

\begin{figure*}[tp]
	\centering
	\includegraphics[width=\textwidth]{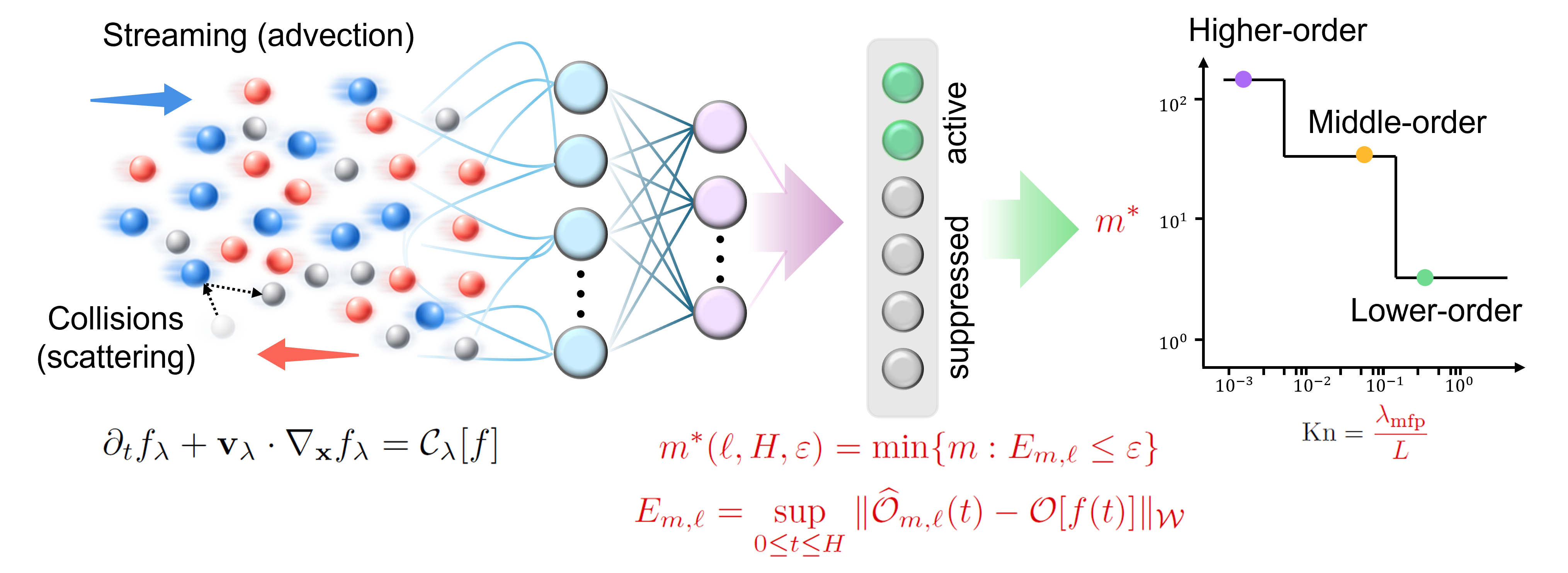}
	\caption{\rev{Predictive state reduction in kinetic transport. Streaming and collisions generate a microscopic distribution, which is encoded into candidate coarse states. State selection retains the information required to predict prescribed observables at spatial resolution $\ell$, horizon $H$, and tolerance $\varepsilon$. The active-state budget is defined by the observable error $E_{m,\ell}$. The staircase illustrates candidate model orders only: its values and its direction with Knudsen number are not numerical results or a universal monotonicity prediction. The subsequent figures test state sufficiency using controlled kinetic preparations.}}
	\label{fig1}
\end{figure*}

Complementary perspectives arise from dynamical systems, observability, and information theory~\cite{LozanoDuran2022,Gross2006,Presse2013}. A microscopic perturbation is relevant to a selected measurement only when it can influence the future measurement trajectory. In a linearized system this influence can be quantified by a finite-horizon observability map, not simply by the decay rate of an isolated mode. In particular, non-normal dynamics and weakly observed relaxation channels can defeat a ranking based on timescales alone~\cite{Asllani2018}. The information-bottleneck principle formalizes the competing demands of compression and retention of information relevant to a target~\cite{Tishby1999}; predictive variants extend this perspective to dynamical variables~\cite{Wang2021}. These ideas suggest defining a coarse state through its \emph{predictive sufficiency}, rather than its instantaneous ability to reconstruct the full microscopic distribution.

Machine learning has provided practical tools for identifying effective coordinates and transport closures. Sparse identification extracts parsimonious dynamical laws from trajectories~\cite{Brunton2016}, while joint representation-and-dynamics learning can reveal low-dimensional coordinates that admit simple evolution equations~\cite{Champion2019}. In molecular kinetics, time-lagged autoencoders and variational networks identify dynamically informative collective variables~\cite{Wehmeyer2018,Mardt2018}. Directly in kinetic theory, learned generalized moments and structure-preserving neural closures have extended reduced modeling across transport regimes while enforcing important physical constraints~\cite{Han2019,Schotthofer2022}. Data-driven Mori--Zwanzig methods additionally learn reduced dynamics with memory~\cite{Lin2021,Lu2026}. These developments establish that learning effective variables and neural kinetic closures are not themselves novel tasks. What remains difficult is to \emph{falsify} a proposed state reduction: to identify microscopic preparations that agree in all retained fields but nevertheless generate different future observables, and to verify that the recovered extra states causally explain the discrepancy.

We therefore formulate state discovery as a test of predictive distinguishability. Consider two physically admissible kinetic preparations with identical resolved fields and the same subsequent forcing. If their future observable trajectories separate, no deterministic closure initialized solely by the common resolved state can predict both exactly. For an ensemble of preparations, the conditional variance of the future trajectory provides a lower bound on the mean-square error of every deterministic predictor based on a given encoded state. In the linearized regime, a finite-horizon observability operator identifies unresolved perturbation directions that produce the largest future observable differences at fixed microscopic amplitude. These results are consequences of established conditional-expectation and observability theory; their role here is to supply explicit, quantitative tests of kinetic closure rather than new general mathematical theorems.
Building on these tests, we propose a scale-conditioned neural reduction that learns a compact state and its dynamics from kinetic trajectories while preserving energy conservation and a prescribed near-equilibrium entropy metric. The effective state dimension is selected by independently validated prediction error, not by latent sparsity alone. A discrete three-velocity model supplies an exactly solvable benchmark: two initial distributions can have identical energy and heat-flux fields yet differ in a higher moment that changes the subsequent energy evolution. The benchmark makes the missing-information problem explicit, and controlled state-removal and response tests specify what would count as successful physical identification.

\begin{figure*}[tp]
	\centering
	\includegraphics[width=\textwidth]{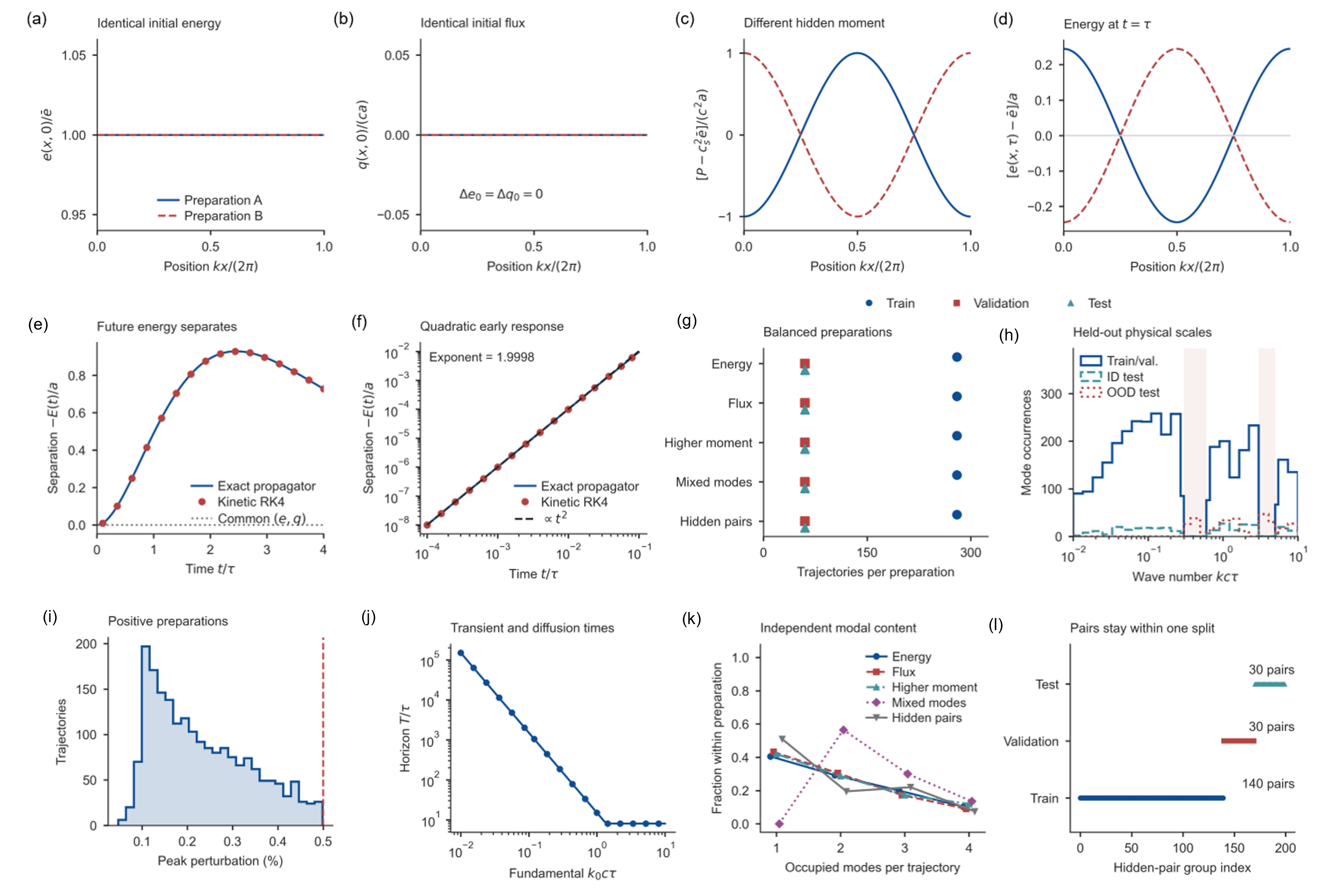}
	\caption{\rev{A hidden-state witness and the three-velocity training ensemble. (a),(b) The two preparations have identical initial energy and heat flux at every position. (c) Their initial second moments differ. (d) The energy profiles separate by $t=\tau$. (e) The energy cosine-amplitude separation $-E(t)/a$ from the exact propagator agrees with an independent kinetic Runge--Kutta calculation. (f) The early response follows $t^2$; fitting $10^{-4}\leq t/\tau\leq10^{-3}$ gives an exponent of $1.9998$. Here $a$ is the amplitude of the pair difference in Eq.~\eqref{eq:hidden-pair}, not of either preparation. (g) Preparation counts by data split. (h) Occupied wave numbers, including held-out bands; ID and OOD denote in-distribution and out-of-distribution tests. (i) Peak population perturbations remain below the indicated positivity bound. (j) Observation horizons resolve both relaxation and long-wavelength diffusion. (k) Distribution of occupied Fourier-mode counts within each preparation family. (l) Each two-member hidden pair remains in one split. The 2000 trajectories are divided into 1400 training, 300 validation, and 300 test samples, with 140, 30, and 30 hidden pairs, respectively. These curves and counts are deterministic; they are not statistical confidence intervals.}}
	\label{fig2}
\end{figure*}

\section{Statistical-mechanical foundations of scale-dependent state reduction}

The emergence of macroscopic transport laws from microscopic statistical dynamics is fundamentally a problem of dynamical state reduction. Although the Boltzmann transport equation (BTE) describes the evolution of mesoscopic one-particle distribution functions, continuum constitutive laws involve substantially fewer dynamical variables. The central question is therefore not merely whether microscopic dynamics can reproduce macroscopic transport, but which microscopic degrees of freedom remain dynamically relevant at a prescribed observation scale and accuracy. We formulate this question through a scale-dependent projection of kinetic dynamics, without assuming a priori that the resulting effective description follows Fourier diffusion, Cattaneo–Vernotte (CV) transport, or any prescribed moment hierarchy.

\begin{revision}
The information flow in Fig.~\ref{fig1} separates three tasks: encoding the microscopic state, selecting an admissible state budget, and testing its future observable response. The schematic does not determine a universal ordering of continuum theories with Knudsen number. The relevant model order must instead be established from the prescribed preparations and observations.
\end{revision}

\subsection{Microscopic kinetic dynamics and macroscopic observables}

Consider a phonon system~\cite{Chen2021} characterized by the distribution function \(f_\lambda(\mathbf x,t)\), where \(\lambda\) denotes the phonon branch and wave vector. In the absence of external forces, its evolution is governed by
\begin{equation}
	\partial_t f_\lambda+
	\mathbf v_\lambda\cdot\nabla_{\mathbf x}f_\lambda
	=
	\mathcal C_\lambda[f],
\end{equation}
where \(\mathbf v_\lambda=\nabla_{\mathbf k}\omega_\lambda\) is the group velocity and \(\mathcal C_\lambda\) is the collision operator. The microscopic energy density and heat flux are obtained from the kinetic moments
\begin{equation}
e(\mathbf x,t)
=
\sum_\lambda
\hbar\omega_\lambda f_\lambda(\mathbf x,t),
\end{equation}
\begin{equation}
\mathbf q(\mathbf x,t)
=
\sum_\lambda
\hbar\omega_\lambda\mathbf v_\lambda
f_\lambda(\mathbf x,t),
\end{equation}
where the summation includes the appropriate wave-vector integration and normalization.
For energy-conserving collisions,
\begin{equation}
\sum_\lambda
\hbar\omega_\lambda\mathcal C_\lambda[f]=0,
\end{equation}
which yields the exact conservation law
\begin{equation} \label{conservation law}
	\partial_t e+\nabla\cdot\mathbf q=0.
\end{equation}

Equation~\eqref{conservation law} is independent of any constitutive approximation. By contrast, a closed evolution equation for \(\mathbf q\) generally requires additional information about the microscopic distribution. This distinction motivates the search for an effective state representation that retains precisely those degrees of freedom required for predictive closure.

We introduce a characteristic length \(L\), observation time \(t_{\mathrm{obs}}\), mean free path $\lambda_{\mathrm{mfp}}$, and microscopic relaxation time \(\tau\). The corresponding dimensionless parameters are
\begin{equation}
\mathrm{Kn}=\frac{\lambda_{\mathrm{mfp}}}{L},
\qquad
\mathrm{De}=\frac{\tau}{t_{\mathrm{obs}}}.
\end{equation}

For a multimode phonon system, these parameters represent characteristic scales rather than a complete description of the relaxation spectrum. The effective dynamics may additionally depend on the distribution of mode-dependent velocities and collision times.

\begin{figure*}[tp]
	\centering
	\includegraphics[width=\textwidth]{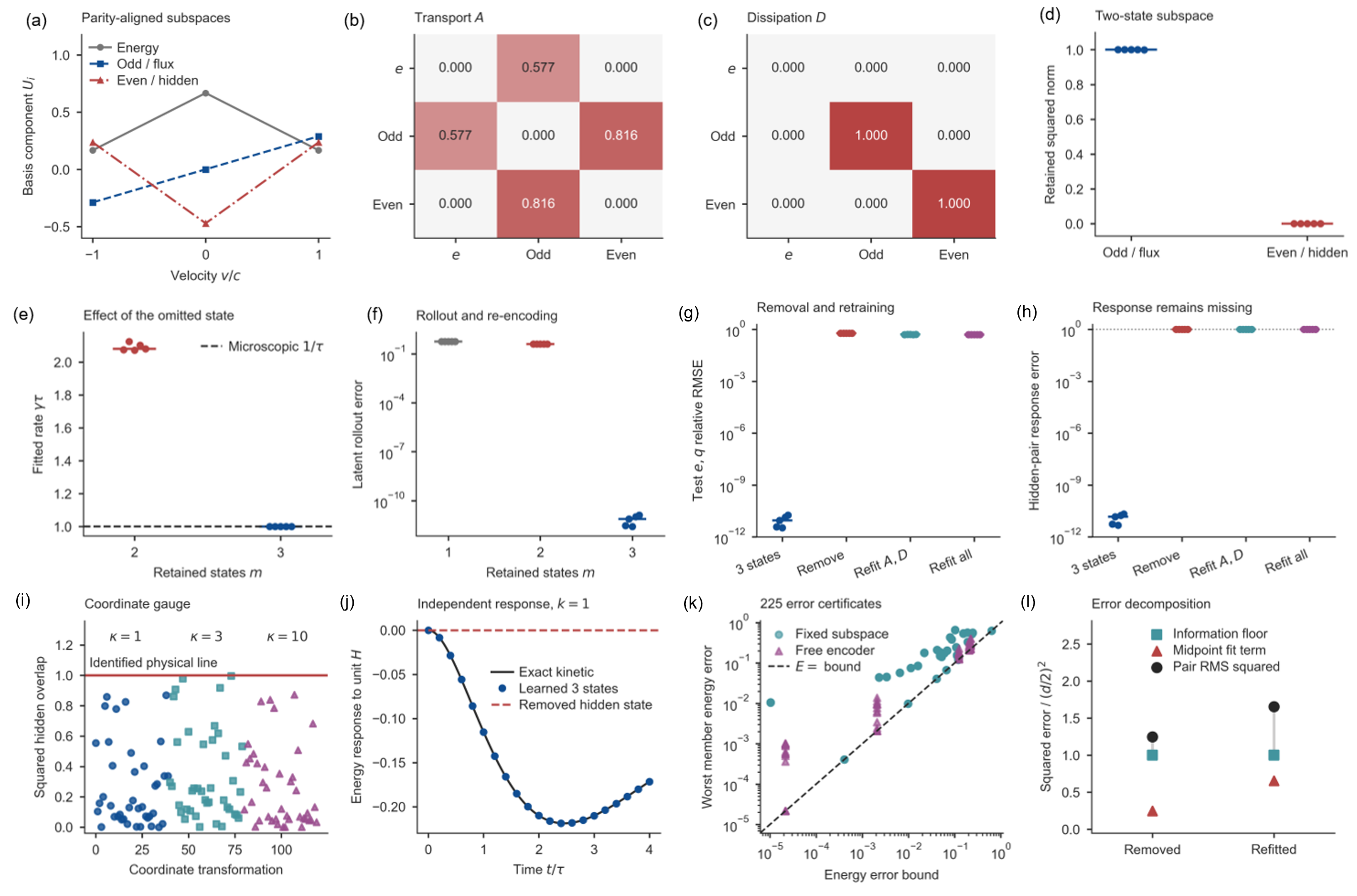}
	\caption{\rev{Recovered three-velocity states and the effect of deleting a predictive direction. (a) Entropy-normalized $m=3$ basis after parity alignment. (b),(c) Transport and dissipation matrices in the aligned basis; the couplings approach $1/\sqrt3$ and $\sqrt{2/3}$ in units $c=\tau=1$. (d) Squared retention of the odd flux and even hidden-moment subspaces by the $m=2$ model. (e) Fitted non-energy damping rates; the dashed line is the microscopic rate $1/\tau$. (f) Latent rollout mismatch relative to re-encoding the exact kinetic state, evaluated at $0.1\tau$, $\tau$, and $5\tau$. (g) Joint energy--flux test errors for the full model, deletion, fixed-subspace dynamics refitting, and refitting of all remaining parameters. (h) Corresponding hidden-pair difference-response errors. (i) Raw-coordinate overlaps under invertible latent transformations, compared with the invariant lifted physical direction. (j) Independent energy response at $kc\tau=1$. (k) Worst-member errors versus matched-input lower bounds for 225 deterministic certificates. (l) Pair-averaged squared RMS error, normalized by $(d/2)^2$, separated into the information floor and midpoint-prediction error. Small points in the seed comparisons denote five optimization seeds; horizontal ticks denote medians. The full three-state basis spans the kinetic space, whereas deleting its hidden direction removes information that refitting alone cannot reconstruct.}}
	\label{fig3}
\end{figure*}

\subsection{Scale-dependent statistical projection}

Let \(f(t)\) denote the full kinetic state in an appropriate function space. We define a scale-dependent coarse-graining map
\begin{equation} \label{eq:coarse-graining map}
	z_\ell(t)=\mathcal P_\ell[f(t)],
\end{equation}
where \(\ell\) denotes the coarse-graining scale (distinct from the mean free path) and \(z_\ell\) is an effective state whose dimension is not prescribed by a conventional moment expansion.
For a linearized kinetic system,
\begin{equation}
\partial_t\delta f
=
\mathcal L_{\mathrm B}\delta f,
\end{equation}
the microscopic evolution is
\begin{equation}
\delta f(t)
=
e^{t\mathcal L_{\mathrm B}}\delta f(0).
\end{equation}

The generator \(\mathcal L_{\mathrm B}\) contains both streaming and collision processes. Its spectral structure determines the relaxation of kinetic perturbations and therefore provides a physical basis for identifying dynamically relevant states.
For an isolated eigenmode,
\begin{equation}
\mathcal L_{\mathrm B}\phi_n
=
\lambda_n\phi_n,
\end{equation}
the associated relaxation time is
\begin{equation}
\tau_n=
\frac{1}{|\operatorname{Re}\lambda_n|},
\end{equation}
provided \(\operatorname{Re}\lambda_n<0\).
Modes with \(\tau_n\ll t_{\mathrm{obs}}\) may become negligible after their initial transients decay, whereas slowly relaxing modes must generally remain represented. Nevertheless, relaxation times alone do not determine the optimal reduction in non-normal systems, where transient amplification and the observability of individual modes can influence their dynamical importance.

The encoder in Eq.~\eqref{eq:coarse-graining map} is a coordinate map and need not be an idempotent projection. Only in the following linear derivation, let \(\mathcal P_\ell\) denote a same-space linear projector with \(\mathcal P_\ell^2=\mathcal P_\ell\). To characterize the reduction more generally, introduce complementary projection operators \(\mathcal P_\ell\) and \(\mathcal Q_\ell=I-\mathcal P_\ell\). For a linear projection, elimination of the unresolved degrees of freedom gives the exact projected evolution
\begin{equation} \label{projected evolution}
	\begin{aligned}
		\frac{d}{dt}\mathcal P_\ell\delta f(t)
		={}&
		\mathcal P_\ell\mathcal L_{\mathrm B}
		\mathcal P_\ell\delta f(t)\\
		&+
		\int_0^t
		\mathcal K_\ell(t-s)
		\mathcal P_\ell\delta f(s)\,ds\\
		&+
		\eta_\ell(t),
	\end{aligned}
\end{equation}
with memory operator
\begin{equation}
\mathcal K_\ell(t)
=
\mathcal P_\ell\mathcal L_{\mathrm B}
e^{t\mathcal Q_\ell\mathcal L_{\mathrm B}}
\mathcal Q_\ell\mathcal L_{\mathrm B}\mathcal P_\ell
\end{equation}
and unresolved initial-state contribution
\begin{equation} \label{initial-state contribution}
\eta_\ell(t)
=
\mathcal P_\ell\mathcal L_{\mathrm B}
e^{t\mathcal Q_\ell\mathcal L_{\mathrm B}}
\mathcal Q_\ell\delta f(0).
\end{equation}

Equations~\eqref{projected evolution}–~\eqref{initial-state contribution} establish an essential constraint on scale-dependent state reduction: eliminating microscopic variables generally generates temporal memory and dependence on unresolved initial conditions. Consequently, a reduced Markovian evolution is justified only when these contributions can be neglected or represented through additional effective variables.

\subsection{Emergence of continuum transport}

In the hydrodynamic limit, rapid collision-induced relaxation drives the distribution toward a local equilibrium manifold. Under suitable assumptions on the collision operator and separation of scales, the leading-order constitutive relation becomes
\begin{equation}
\mathbf q=-\kappa\nabla T,
\end{equation}
which, together with Eq.~\eqref{conservation law} and $\delta e\simeq C\delta T$, gives
\begin{equation}
	\partial_tT=\alpha\nabla^2T,
	\qquad
	\alpha=\frac{\kappa}{C}.
\end{equation}

A description retaining an additional heat-flux relaxation variable may instead take the form
\begin{equation}
\tau_q\partial_t\mathbf q+\mathbf q
=
-\kappa\nabla T,
\end{equation}
leading to
\begin{equation} \label{eq:CV}
	\tau_q\partial_t^2T+
	\partial_tT
	=
	\alpha\nabla^2T.
\end{equation}

Equation~\eqref{eq:CV} represents a possible finite-dimensional closure~\cite{Guyer1966} rather than a universal intermediate limit of the BTE. In particular, multiple relaxation channels and spatially nonlocal effects can require higher-dimensional or non-Markovian constitutive descriptions.
These observations suggest that the number of effective dynamical variables should be determined by predictive sufficiency rather than by an assumed hierarchy of macroscopic equations.

\subsection{Minimal predictive state and dynamical complexity}

We define a scale-dependent effective state
\begin{equation}
\mathbf z_\ell
=
(z_1,\ldots,z_m)
\end{equation}
as predictively sufficient over a specified horizon \(H\) when its evolution reproduces a prescribed set of observables within tolerance \(\varepsilon\).

Let \(\mathcal O[f]\) denote the observable map, and let \(\widehat{\mathcal O}_{m,\ell}(t)\) be the prediction obtained from an \(m\)-dimensional reduced state. The minimal predictive dimension is defined operationally as

\begin{equation} \label{minimal predictive dimension}
\begin{aligned}
 m^*(\ell,H,\varepsilon)&=\min\{m:E_{m,\ell}\leq\varepsilon\},\\
 E_{m,\ell}&=\inf_{\mathcal M\in\mathfrak A_{m,\ell}}
 \sup_{f_0\in\mathcal I}\sup_{0\leq t\leq H}\\
 &\quad\|\widehat{\mathcal O}_{\mathcal M}(t;f_0)
 -\mathcal O[f(t;f_0)]\|_{\mathcal W}.
\end{aligned}
\end{equation}

Here $\mathfrak A_{m,\ell}$ is the declared admissible model class, $\mathcal I$ is the preparation set, and each candidate uses its specified encoder to initialize from $f_0$. The infimum defines ideal approximation error; existence of a minimizing model is not assumed. The error norm \(\|\cdot\|_{\mathcal W}\) incorporates the relative importance of the selected observables. The definition is understood over a specified class of initial conditions and excitations, with the approximation required to generalize beyond the training trajectories.
The supremum norm in Eq.~\eqref{minimal predictive dimension} is a theoretical sufficiency criterion. The numerical dimension maps below use a continuous-window, fixed-scale RMS criterion and a declared finite model bank. We denote their result by $\widehat m^*_{\mathrm{bank}}$; an RMS pass does not certify the supremum criterion or a global minimum over all autonomous closures.

Equation~\eqref{minimal predictive dimension} provides a measurable characterization of dynamical complexity. Unlike a fixed moment expansion, it permits the effective state dimension to vary with spatial scale, temporal resolution, material properties, and predictive accuracy.
The resulting theoretical hypothesis is that kinetic-to-continuum transport can be characterized by a scale-dependent reduction of predictive state complexity,
\begin{equation} \label{scale-dependent reduction}
	m^*
	=
	m^*(\mathrm{Kn},\mathrm{De},H,\varepsilon),
\end{equation}
where changes in \(m^*\) reflect the progressive loss of dynamically relevant microscopic information. Such changes need not constitute sharp physical transitions; rather, they identify scales at which a reduced description becomes sufficient under a specified error criterion.

All minimizations are relative to a specified admissible model class and preparation set. In a worst-case formulation the error must also take a supremum over that set, and the model parameters must be optimized within each rank. We distinguish this ideal criterion from a passing finite validation sample. Appendix~\ref{app:closure} gives an exact linear closure condition and a quantitative consistency bound; Appendix~\ref{app:bounds} specifies the admissible perturbation set for information bounds. The general nonlinear architecture is retained in Appendix~\ref{sec:architecture}; the main text develops the linear realization actually tested.
The objective is therefore to discover both the effective state space and its evolution law directly from microscopic trajectories, while retaining the conservation and thermodynamic structure of the underlying kinetic dynamics.

\begin{figure*}[tp]
	\centering
	\begin{overpic}[width=\textwidth]{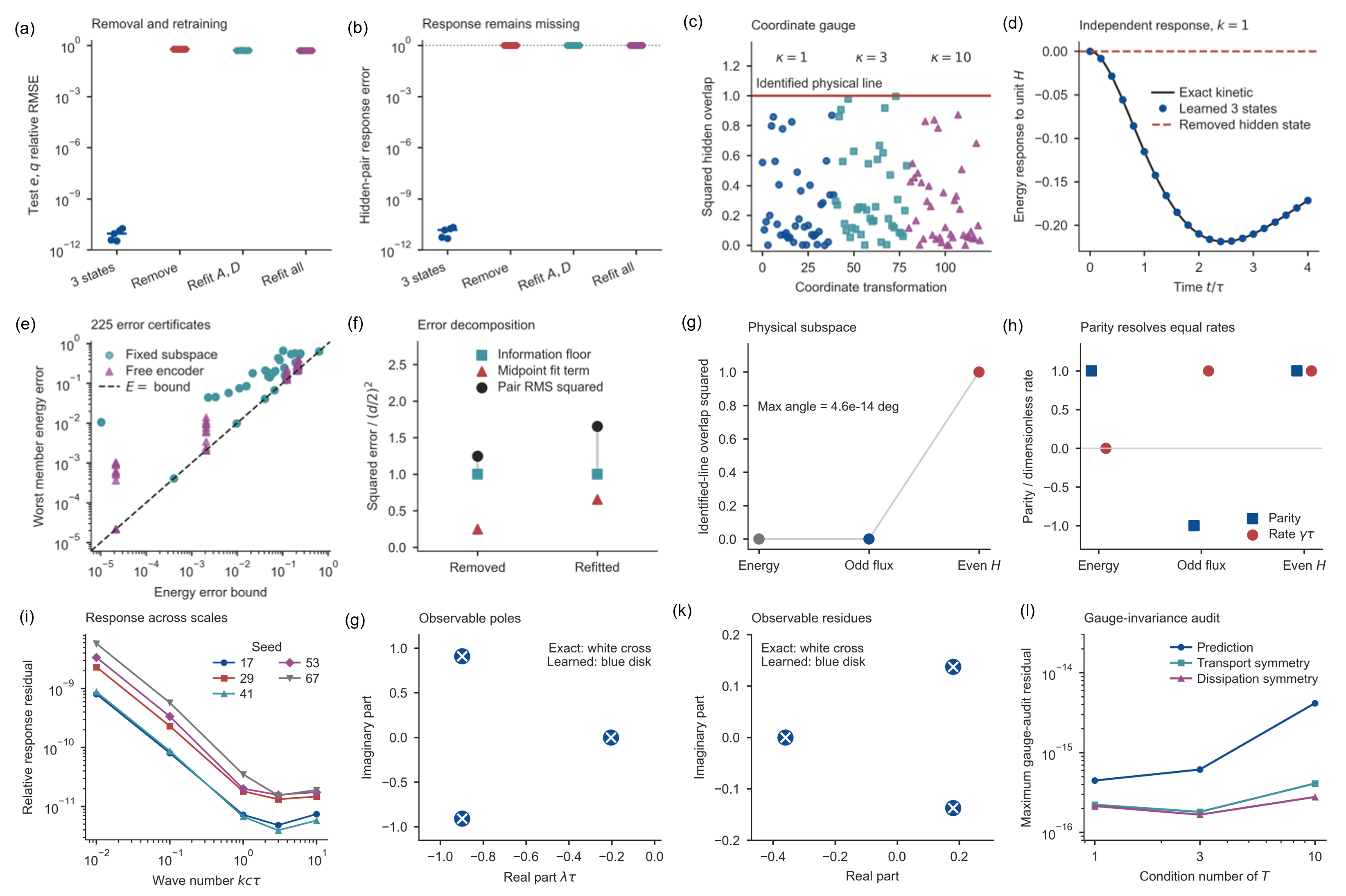}
		\put(25.25,19.5){\color{white}\rule{0.033\textwidth}{0.025\textwidth}}
		\put(25.65,20.0){\color{black}\sffamily\fontsize{7}{8}\selectfont(j)}
	\end{overpic}
	\caption{Separating missing information from fitting error. (a),(b) Joint-observable and hidden-pair response errors before deletion, after deletion, after fixed-subspace dynamics refitting, and after free-encoder refitting; points and ticks denote five seeds and their medians. (c) Coordinate overlaps vary under invertible transformations with condition numbers $\kappa=1,3,10$, while the lifted hidden subspace remains unchanged. (d) An independently initialized hidden perturbation distinguishes the full and deleted models. (e) Actual worst-member energy errors for 225 matched-input certificates; the dashed line is equality with the corresponding lower bound, including the finite latent-matching correction. (f) The normalized squared pair RMS error is the sum of the information floor and midpoint-fit term for a representative $kc\tau=1$ pair. (g) Squared overlaps of the training-identified physical line with energy, odd flux, and the even hidden direction $H_{\mathrm{hid}}$, denoted $H$ in the panel. (h) Velocity parity and collision rate distinguish the two equally relaxing non-energy directions. (i) Independent response residuals over wave number for all five seeds. (j),(k) Observable complex poles and residues, with exact values marked by crosses. (l) Maximum residuals of prediction, transport symmetry, and dissipation symmetry under the coordinate transformations. For a nonorthogonal transformation, the last two tests use the transformed metric, as verified in Appendix~\ref{app:reporting} and defined in Appendix~\ref{app:physical}. Certificates and transformations are deterministic diagnostics, not independent physical realizations.}
	\label{fig4}
\end{figure*}

\subsection{Hidden information as a failure of predictive sufficiency}
Fix an initial preparation ensemble, material parameters, boundary forcing, spatial filter \(\ell\), sampling interval \(\Delta t\), and an observation window \([t_w,t_w+H]\). Let \(X\) be the kinetic state at initialization and
\begin{equation}
 Y=\big(\mathcal O_\ell[f(t_w+t_j)]\big)_{j=1}^{N_t},
 \qquad Z=\mathcal E_\theta(X).
\end{equation}
The norm of \(Y\) uses prescribed time, spatial-quadrature, and observable weights; each observable is normalized by a fixed physical scale. For deterministic kinetic evolution with fixed future forcing, \(Y\) is determined by \(X\). For any deterministic predictor \(h(Z)\), conditional expectation gives
\begin{equation}\label{eq:variance}
\begin{aligned}
\mathbb E\|Y-h(Z)\|_W^2
={}&\mathbb E\|Y-\mathbb E[Y|Z]\|_W^2\\
&+\mathbb E\|\mathbb E[Y|Z]-h(Z)\|_W^2.
\end{aligned}
\end{equation}
The cross term vanishes because its conditional expectation at fixed \(Z\) is zero. Thus
\begin{equation}\label{eq:V}
 V(Z)=\mathbb E\,\mathrm{tr}\!\left[W\,\mathrm{Cov}(Y|Z)\right]
\end{equation}
is the irreducible mean-square error for this representation, even with an ideal predictor. A small training error does not establish that \(V(Z)\) is small outside the training ensemble.

Draw two preparations independently from the same conditional ensemble \(X,X'|Z=z\). Expanding their squared future separation gives the exact identity
\begin{equation}\label{eq:pair}
\frac12\mathbb E\!\left[\|Y-Y'\|_W^2\,\middle|\,Z=z\right]
=\mathrm{tr}\!\left[W\,\mathrm{Cov}(Y|Z=z)\right].
\end{equation}
This connects a statistical-mechanical ensemble of unresolved states to a directly measurable closure defect. Exact conditional sampling is generally difficult. Arbitrary adversarial pairs do \emph{not} estimate Eq.~\eqref{eq:pair}; instead, if their encoded states are identical and their future separation is \(d\), the triangle inequality proves that any common deterministic prediction has error at least \(d/2\) for one member. Approximate latent matches introduce bias and must report their matching tolerance. If the predictor is \(L_h\)-Lipschitz, the corresponding lower bound is \(\max[0,d-L_h\|Z-Z'\|]/2\).

For stochastic dynamics, \(V(Z)\) also contains intrinsic process noise. Define the excess predictable variance
\begin{equation}
 V_{\mathrm{hid}}(Z)=\mathbb E\|\mathbb E[Y|X]-\mathbb E[Y|Z]\|_W^2
\end{equation}
for deterministic \(Z=\mathcal E_\theta(X)\). The law of total variance separates this term from \(\mathbb E\,\mathrm{tr}[W\mathrm{Cov}(Y|X)]\); replicated stochastic simulations are needed to estimate the latter. Hidden information cannot be equated with all observed noise.

\subsection{A microscopic witness for missing macroscopic variables}
Let \(B X\) collect the resolved fields, initially energy alone or energy and heat flux. Construct physically admissible pairs
\begin{equation}\label{eq:preparations}
 X_\pm=X_0\pm a h,\qquad Bh=0,
\end{equation}
with identical material and forcing. Positivity of the occupations restricts \(a\). The full resolved spatial fields, not just their domain averages, must match. In linearized dynamics, set
\begin{equation}\label{eq:obs}
 \mathsf O_H h=
 \begin{pmatrix}
 W_1^{1/2}B_o e^{t_1\mathsf L}h\\
 \vdots\\
 W_{N_t}^{1/2}B_o e^{t_{N_t}\mathsf L}h
 \end{pmatrix},
\end{equation}
where \(B_o\) is the future observation map and the times include the chosen waiting time. If \(Bh=0\) but \(\mathsf O_Hh\ne0\), then the present resolved fields are insufficient to determine their measured future. Their pair separation is \(2|a|\|\mathsf O_Hh\|\).

Choose a positive-definite microscopic metric \(G\), preferably the negative entropy Hessian at equilibrium, and a basis \(N\) with
\begin{equation}
 BN=0,\qquad N^\dagger GN=I.
\end{equation}
The singular values of \(\mathsf O_HN\) rank unresolved perturbations by their future observable effect at equal thermodynamic amplitude. Equivalently,
\begin{equation}\label{eq:gram}
 (\mathsf O_HN)^\dagger(\mathsf O_HN)
 =N^\dagger\sum_j e^{t_j\mathsf L^\dagger}
 B_o^\dagger W_j B_o e^{t_j\mathsf L}N.
\end{equation}
For linear encoders restricted to this unresolved subspace and arbitrary linear future decoders, the best rank-\(r\) operator approximation has worst-case unit-amplitude error \(\sigma_{r+1}(\mathsf O_HN)\). Hence an error tolerance \(\varepsilon\) at amplitude \(a\) requires
\begin{equation}\label{eq:rank}
 r\geq\#\{i: |a|\sigma_i(\mathsf O_HN)>\varepsilon\}.
\end{equation}
This is a finite-horizon linear representation bound, not a sufficiency theorem for autonomous nonlinear closure. Semigroup consistency and rollout tests remain necessary. For a large kinetic system, matrix-free tangent and adjoint propagations can estimate leading directions without assembling the full operator.

\subsection{What counts as a physical discovery}
A recovered latent subspace is interpretable when microscopic perturbations along it change the predicted observables, removing it destroys that prediction, and restoring it repairs the failure on independent preparations. Local encoder gradients can be compared, in the \(G\) metric, with energy, odd velocity moments, even higher moments, and collision eigenfunctions. Report subspace angles, parity, relaxation rates, and response residues; correlations with a single latent coordinate are gauge dependent. If only temperature snapshots are available experimentally, arbitrary hidden kinetic states are not identifiable from one snapshot. A history encoder may help, but it solves a different observation problem.

The arguments of Eq.~\eqref{scale-dependent reduction} must therefore include the preparation ensemble, observation map, \(t_w\), \(\Delta t\), and state-memory budget. Here \(m\) counts fields per coarse spatial cell, not the number of grid points. With arbitrary initial fluxes and observations beginning at \(t_w=0\), a temperature-only Fourier closure cannot achieve arbitrarily small error. Apparent dimension reduction can instead follow removal of an initial transient, reduced measurement bandwidth, restricted preparations, or genuine scale separation. These possibilities must be distinguished.

\section{Structure-preserving learned reduction and analytical benchmark}\label{sec:realization}
\subsection{Minimal implementable model}\label{sec:linear_realization}
The demonstrated realization uses one-dimensional, near-equilibrium discrete-velocity transport, with periodic boundaries and a fixed reference temperature. This isolates variable discovery from uncertain material parameters, nonlinear collisions, and boundary layers. Let \(x\in\mathbb R^{N_v}\) be the local kinetic perturbation and \(G\) its fixed positive entropy metric. Use the learnable projection
\begin{equation}\label{eq:encoder}
 z=U^TGx,\qquad U^TGU=I,\qquad \widehat x=Uz.
\end{equation}
One basis vector is fixed along \(G^{-1}b_e\), where \(b_e^Tx\) is energy; all other vectors are learned without identifying them as flux or higher moments. A network conditioned on physical scales outputs the free columns of \(U\), followed by orthonormalization in the \(G\) metric. Anchoring the known conservation law removes an unidentifiable coordinate freedom; it does not prescribe the missing constitutive variables. A fully unconstrained encoder is a useful ablation, not a requirement for discovery.

For \(\partial_tx=-V\partial_xx+C_Bx\), suppose \(GV\) is symmetric and \(GC_B\) is symmetric negative semidefinite, as in a reversible linearized collision model. Then
\begin{equation}\label{eq:galerkin}
 \partial_tz+A\partial_xz=-Dz,\quad
 A=U^TGVU,\quad D=-U^TGC_BU
\end{equation}
is a learnable projected dynamics with \(A=A^T\) and \(D\succeq0\). Learn \(U\) through trajectory losses and differentiate through the propagator. This restricted linear model is the essential transparent baseline, even if the scale-to-basis map is neural. If the collision operator is not available, fit \(A=(K+K^T)/2\) and \(D=\Pi RR^T\Pi\), with \(\Pi=I-e_1e_1^T\). Enforce the energy-flux row against the microscopic flux observable. The energy component has no collision source; a conservative spatial discretization then preserves total energy exactly.

For periodic boundaries,
\begin{equation}\label{eq:entropy}
 \frac{d}{dt}\frac12\int z^Tz\,dx
 =-\int z^TDz\,dx\leq0.
\end{equation}
This is decay of the quadratic entropy deficit, not decay of physical thermal energy. For phonon occupations at \(f^0_\lambda>0\), the microscopic entropy is
\begin{equation}
 S=k_B\sum_\lambda w_\lambda
 [(1+f_\lambda)\ln(1+f_\lambda)-f_\lambda\ln f_\lambda],
\end{equation}
so \(G_{\lambda\lambda}=k_Bw_\lambda/[f^0_\lambda(1+f^0_\lambda)]\) is its negative Hessian. Equation~\eqref{eq:entropy} applies to the small-perturbation regime and the stated collision assumptions. A nonlinear extension requires a realizable distribution decoder and a compatible kinetic entropy; it cannot inherit this proof merely by adding nonlinear layers.

Use ranks \(m=1,2,3,4,6,8\) where available, training each rank separately or using hard pruning followed by refitting. Soft gates alone are inadequate because an encoder can rescale a weakly gated channel. Fix the latent normalization, constrain operator norms, and evaluate the exported hard-pruned model. A practical conditioner is a two-hidden-layer MLP of width 64 with smooth activations and dimensionless scale inputs. At fixed physical parameters and resolution, the generator must not depend on the requested forecast horizon. The horizon selects model order; it is not a new material parameter.

The energy-only Galerkin projection is not the Fourier baseline: in a velocity-symmetric equilibrium it has $A=D=0$ and hence no diffusion. Fourier diffusion requires elimination of relaxing variables or an independently fitted gradient closure. Its algebraic flux decoder also uses a spatial derivative. Thus model-bank comparisons count fields while explicitly allowing different spatial differential orders; they do not compare identical generator classes. Appendix~\ref{app:diffusion} derives the distinction. Thermodynamic contraction alone does not imply positivity of reconstructed populations (Appendix~\ref{app:physical}).

\begin{figure*}[tp]
	\centering
	\includegraphics[width=\textwidth]{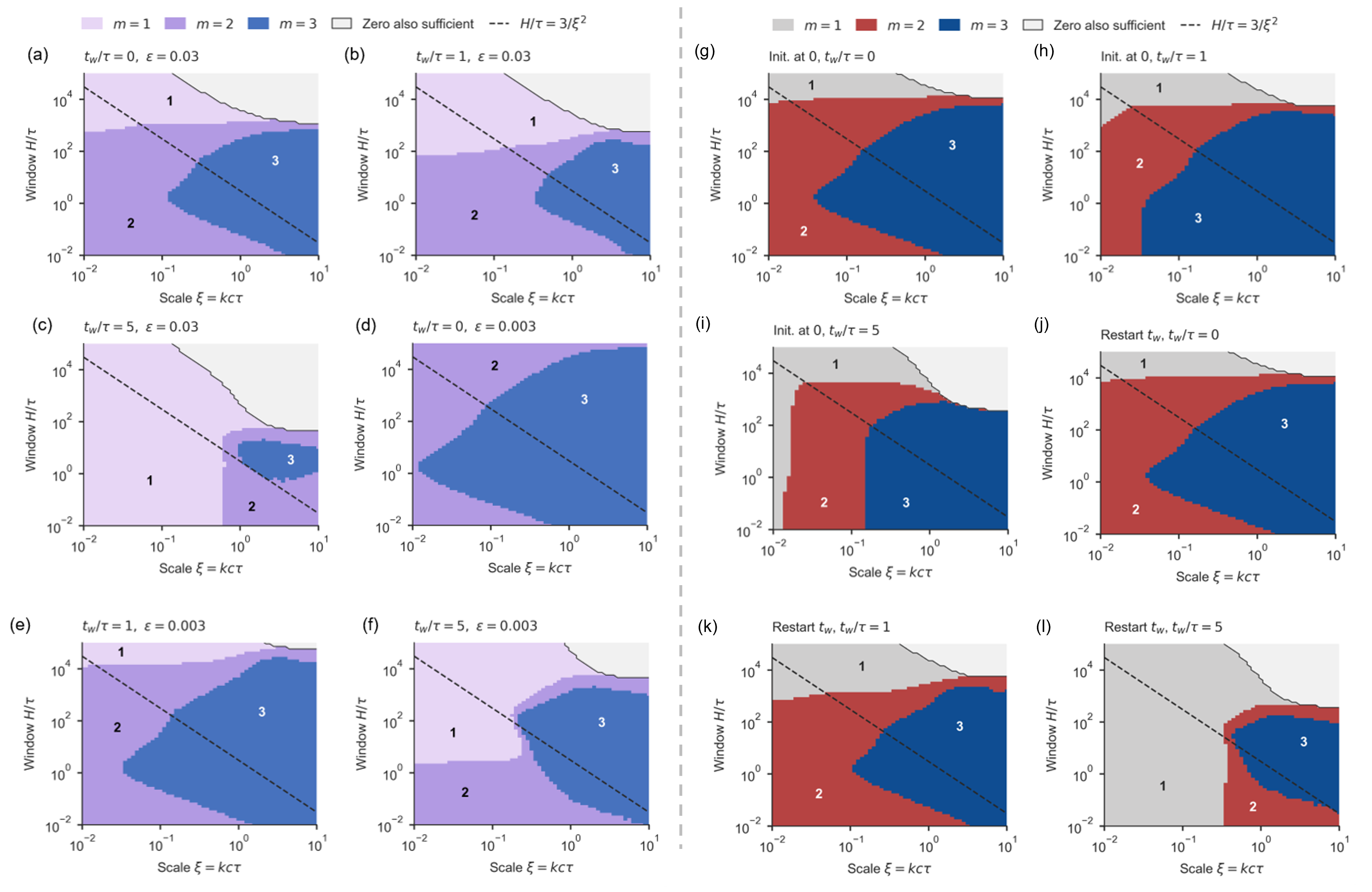}
	\caption{\rev{Predictive dimension depends on the observation and initialization protocol. Colors and state numbers identify the smallest passing model in the declared one-, two-, and three-state bank for joint energy--flux observations. (a)--(c) Re-encoding at the observation start, with tolerance $\varepsilon=0.03$ and waiting times $t_w/\tau=0,1,5$, respectively. (d)--(f) The same waiting times at $\varepsilon=0.003$. (g)--(i) Encoding at $t=0$ followed by delayed observation, at $\varepsilon=0.01$ and $t_w/\tau=0,1,5$. (j)--(l) Re-encoding the exact kinetic state at those waiting times, at the same tolerance. Each block has its own color key; state numbers, rather than hue alone, identify equal dimensions across the blocks. The pale region also admits the zero predictor and therefore establishes no necessary nonzero state budget. Dashed lines indicate the long-wavelength diffusion time $H_D/\tau=3/\xi^2$, where $\xi=kc\tau$; they are not transition boundaries. Errors are continuous-window RMS values on $[t_w,t_w+H]$, normalized by fixed initial entropy amplitude and output scales. The initial perturbation ball is held fixed between protocols. The maps report achieved model-bank dimensions, not proven global minimum Markov dimensions.}}
	\label{fig5}
\end{figure*}

\subsection{An exact three-velocity witness}\label{sec:three_witness}
Consider velocities \(v_i\in\{-c,0,c\}\), symmetric positive weights \(w_+=w_-\), and \(\sum_iw_i=1\). A positive energy-population model satisfies
\begin{equation}\label{eq:three}
 \partial_tf_i+v_i\partial_xf_i
 =-\frac{f_i-w_i e}{\tau},\qquad e=\sum_i f_i.
\end{equation}
This is a controlled kinetic benchmark, not a complete material-specific phonon BTE. Define
\begin{equation}
 q=\sum_i v_if_i,\quad P=\sum_i v_i^2f_i,\quad
 c_s^2=\sum_iw_iv_i^2.
\end{equation}
Taking moments and using \(v_i^3=c^2v_i\) gives an \emph{exact} closed system:
\begin{equation}\label{eq:three-moments}
\begin{aligned}
 \partial_te+\partial_xq&=0,\\
 \partial_tq+\partial_xP&=-q/\tau,\\
 \partial_tP+c^2\partial_xq&=-(P-c_s^2e)/\tau.
\end{aligned}
\end{equation}
Let \(r=P-c_s^2e\). Its equation is
\begin{equation}
 \partial_tr+(c^2-c_s^2)\partial_xq=-r/\tau.
\end{equation}
Thus \(r\) is a candidate hidden state whose need must be discovered rather than supplied to the network.

Prepare two states whose difference is
\begin{equation}\label{eq:hidden-pair}
 (\Delta f_+,\Delta f_0,\Delta f_-)
 =a\cos(kx)(1,-2,1).
\end{equation}
For a sufficiently small \(a\), both preparations remain positive around a common equilibrium. They satisfy \(\Delta e=\Delta q=0\) everywhere, but \(\Delta P=2c^2a\cos(kx)\). Consequently,
\begin{equation}\label{eq:short}
 \Delta e(x,t)=-c^2ak^2t^2\cos(kx)+O(t^3).
\end{equation}
Indeed, the first time derivative vanishes and \(\partial_t^2\Delta e|_0=\partial_x^2\Delta P|_0\). No deterministic closure initialized only by the common \((e,q)\) fields can reproduce both futures. Adding spatial derivatives of those identical fields cannot repair this failure. This provides a decisive test against a network that has learned only correlations within equilibrium-like preparations.

For a single nonzero Fourier mode, the generator in \((e,q,P)\) coordinates is
\begin{equation}
 L(k)=\begin{pmatrix}
 0&-ik&0\\
 0&-1/\tau&-ik\\
 c_s^2/\tau&-ikc^2&-1/\tau
 \end{pmatrix}.
\end{equation}
With energy observation \(B_e=(1,0,0)\),
\begin{equation}\label{eq:observability}
 \begin{pmatrix}B_e\\B_eL\\B_eL^2\end{pmatrix}
 =\begin{pmatrix}1&0&0\\0&-ik&0\\0&ik/\tau&-k^2\end{pmatrix}
\end{equation}
has determinant \(ik^3\ne0\). All three coordinates are observable from an ideal time-resolved energy trajectory at \(k\ne0\); an exact linear Markov realization for arbitrary preparations requires three coordinates per complex Fourier amplitude. This statement does not identify three real global degrees of freedom for a whole spatial field. At \(k=0\), energy alone is constant and the other coordinates are invisible to that observation.

Neglecting \(r\) only where its feedback is small gives \(P\simeq c_s^2e\), followed by
\begin{equation}
 \tau\partial_t^2e+\partial_te
 =\tau c_s^2\partial_x^2e.
\end{equation}
Neglecting flux relaxation on still longer times gives diffusivity \(\alpha=\tau c_s^2\). These are asymptotic approximations, not an exact three-to-two-to-one progression at arbitrary wave number. In particular, this single-time model need not have a broad CV regime. A separate two-velocity model with velocities \(\pm c\) and weights \(1/2\) has \(P=c^2e\) identically and supplies an exact CV positive control. A model that invents a necessary third state for that control fails the dimension test.

\subsection{Memory is retained information in another representation}
Solving the hidden-moment equation yields
\begin{equation}
\begin{aligned}
 r(t)={}&e^{-t/\tau}r(0)\\
 &-(c^2-c_s^2)\int_0^t e^{-(t-s)/\tau}\partial_xq(s)\,ds.
\end{aligned}
\end{equation}
Substitution into the flux equation gives
\begin{equation}\label{eq:exact-memory}
\begin{aligned}
 \partial_tq={}&-q/\tau-c_s^2\partial_xe
 -e^{-t/\tau}\partial_xr(0)\\
 &+(c^2-c_s^2)\int_0^t e^{-(t-s)/\tau}\partial_x^2q(s)\,ds.
\end{aligned}
\end{equation}
The sign, spatial derivative, and initial-slip term follow directly from moment elimination. A memory model initialized with zero history and no information about \(r(0)\) still cannot distinguish the paired preparations. Report either the total augmented Markov dimension \(m+n_{\mathrm{aux}}\), or the pair \((m,K)\) for \(K\) stored history samples at a stated \(\Delta t\); do not claim that an infinite-memory scalar is a one-state Markov model.

\section{Numerical tests of predictive state sufficiency}\label{sec:numerical_results}
The calculations in Figs.~\ref{fig2}--\ref{fig8} test the linear realization of Sec.~\ref{sec:linear_realization}. The energy coordinate is anchored by conservation, while the remaining encoder directions are learned without assigning them to heat flux or a prescribed higher moment. The scale-conditioned nonlinear architecture of Sec.~\ref{sec:architecture} provides a broader framework; the present evidence concerns independently trained finite-rank linear models and controlled kinetic responses.

\subsection{Matched initial observables, distinct kinetic futures}
Figure~\ref{fig2}(a)--(f) realizes the analytical witness of Sec.~\ref{sec:three_witness}. The initial energy and heat-flux fields coincide pointwise, while the second moments differ. At $t=\tau$, the energy fields have separated, although both preparations had the same resolved $(e,q)$ state. Agreement between the matrix-exponential reference and independent kinetic time integration establishes that this separation is a dynamical response rather than a time-stepping artifact. The fitted early-time exponent $1.9998$ agrees with Eq.~\eqref{eq:short} and identifies the hidden moment's leading feedback on the measured energy.

The accompanying data audit, Fig.~\ref{fig2}(g)--(l), prevents this test from becoming an accidental correlation in the preparation ensemble. The 2000 trajectories include independent energy, flux, higher-moment, mixed-mode, and hidden-pair preparations. Complete preparations are assigned to the 70/15/15 split before fitting, and both members of each hidden pair remain together. Held-out wave-number bands test transfer in physical scale, while the longer observation horizons retain the slow diffusive response. Thus the learner must predict independent microscopic perturbations, rather than infer a hidden moment from its correlation with energy in equilibrium-like data.

\subsection{Recovery and removal of the three-velocity hidden state}
The three-state encoder recovers the complete kinetic subspace. After energy anchoring and parity alignment, Fig.~\ref{fig3}(a)--(c) separates the odd flux direction from an even non-energy direction and recovers the transport couplings $1/\sqrt3$ and $\sqrt{2/3}$, with unit non-energy relaxation in $c=\tau=1$ units. This interpretation is made after training by comparing subspaces; it is not an assignment of a particular raw latent coordinate to $P$. By contrast, the two-state model predominantly retains the odd direction [Fig.~\ref{fig3}(d)]. Its fitted damping exceeds the microscopic rate [Fig.~\ref{fig3}(e)], consistent with partial compensation for omitted hidden-state feedback. That compensation does not produce an autonomous representation: the latent rollout remains inconsistent with microscopic re-encoding [Fig.~\ref{fig3}(f)].

Deletion makes the missing information operational. Removing the identified hidden direction raises both joint-observable and pair-response errors, and refitting the retained dynamics or the remaining encoder does not restore the full-rank accuracy [Fig.~\ref{fig3}(g),(h)]. An independently initialized hidden perturbation produces a nonzero energy response that the deleted model misses [Fig.~\ref{fig3}(j)]. The gauge and certificate tests in Fig.~\ref{fig3}(i),(k),(l) connect this failure to the retained information, rather than to the name or normalization of one latent variable.

\subsection{Information certificates and coordinate-invariant identification}
Figure~\ref{fig4} resolves the complementary diagnostics underlying this conclusion. For a pair with common encoded input, let $Y_+$ and $Y_-$ be its two exact future trajectories, $\overline Y=(Y_++Y_-)/2$, and $d=\|Y_+-Y_-\|_W$. A common deterministic prediction $\widehat Y$ obeys
\begin{equation}\label{eq:finite_pair_decomposition}
\frac{\|Y_+-\widehat Y\|_W^2+\|Y_--\widehat Y\|_W^2}{2}
=\frac{d^2}{4}+\|\overline Y-\widehat Y\|_W^2.
\end{equation}
This finite-pair identity separates an information floor from the error in fitting the pair midpoint; it does not require the adversarial pairs to sample a conditional ensemble. All 225 matched-input certificates respect the corresponding worst-member bound [Fig.~\ref{fig4}(e)], and the representative error decomposition in Fig.~\ref{fig4}(f) exposes both contributions. Global refitting need not reduce the midpoint error of every individual pair.

The physical line identified from training pairs aligns with the even hidden-moment subspace and has negligible overlap with energy or odd flux [Fig.~\ref{fig4}(g)]. Flux and hidden moment share the same collision rate, so their parity, independent response, and observable residues are needed to distinguish them [Fig.~\ref{fig4}(h)--(k)]. Invertible changes of latent coordinates strongly alter raw-coordinate overlaps, while the lifted physical line and its predictions remain invariant to numerical precision [Fig.~\ref{fig4}(c),(l)]. For nonorthogonal coordinate changes, symmetry and norms must be tested in the transformed metric, not in the Euclidean metric (Appendix~\ref{app:physical}). Together with deletion and refitting [Fig.~\ref{fig4}(a),(b)], these tests support the necessity of a predictive subspace, rather than a privileged latent coordinate.

\subsection{Observation-dependent state budgets and the gap to autonomous closure}
Figure~\ref{fig5} evaluates $\widehat m^*_{\mathrm{bank}}$ over $\xi=kc\tau$, forecast horizon, waiting time, and tolerance for joint energy--flux observations. The bank contains a fitted Fourier model with an algebraic flux decoder, a fitted rank-two model, and the learned rank-three model. Tightening the tolerance expands the region requiring three states [Fig.~\ref{fig5}(a)--(f)]. Delaying observation can reduce the required budget after fast transients decay, but the result also depends on whether the reduced model was initialized at zero or re-encoded from the exact kinetic state at $t_w$ [Fig.~\ref{fig5}(g)--(l)]. Those protocols provide different initialization information. Pale regions, where even a zero prediction satisfies the fixed-scale RMS tolerance, indicate insufficient observable signal to establish a nonzero necessary budget. The dashed diffusion-time reference does not define a transition between constitutive laws.

\begin{figure*}[tp]
\centering
\includegraphics[width=\textwidth]{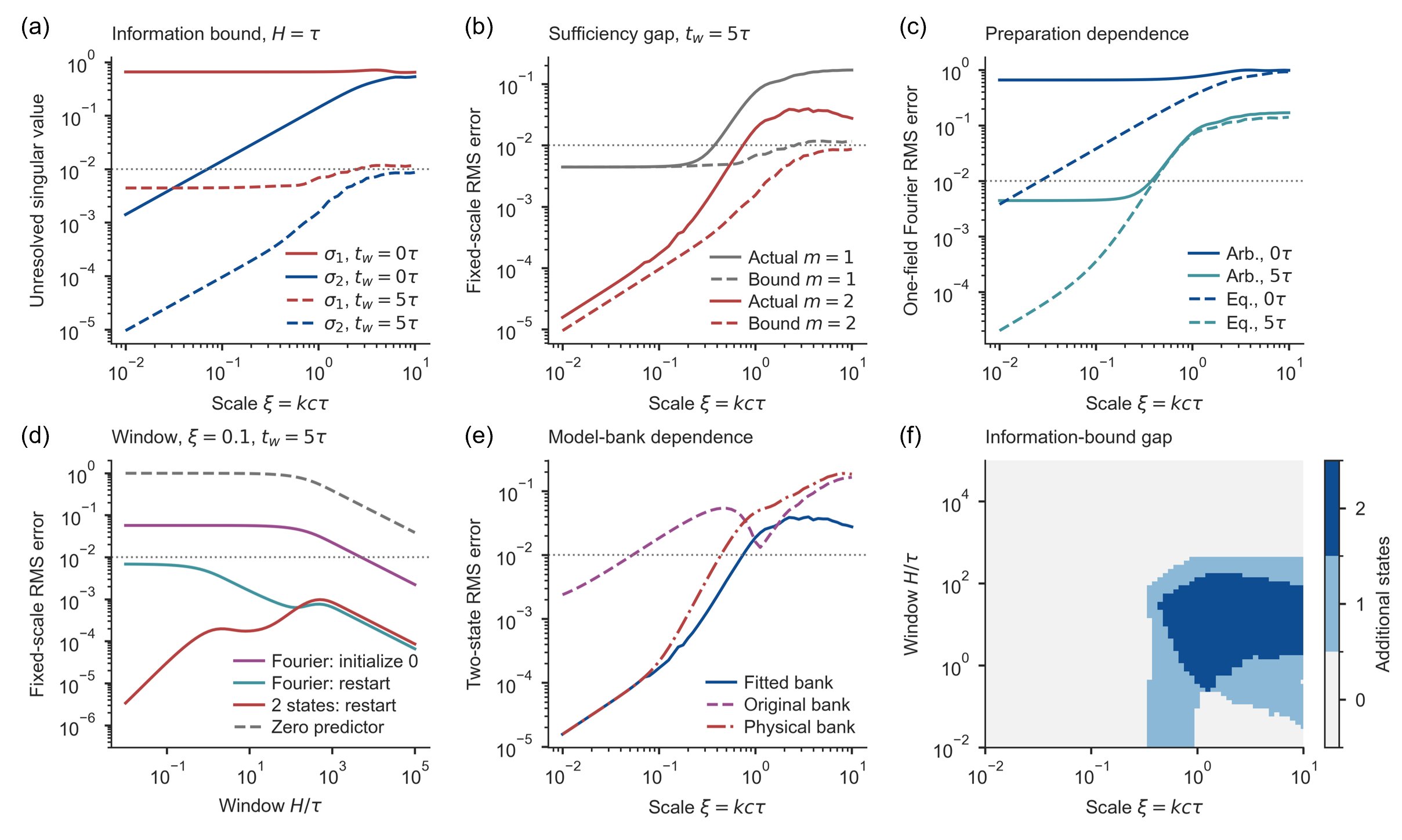}
\caption{\rev{Information bounds, preparation dependence, and model-bank limitations. (a) Singular values of the future-response map restricted to perturbations with zero current energy, for horizon $H=\tau$ and waiting times $t_w/\tau=0,5$. (b) Actual one- and two-state RMS errors and linear representation lower bounds at $t_w=5\tau$, $H=\tau$. (c) One-field Fourier errors for arbitrary preparations (Arb.) and equilibrium energy perturbations (Eq.), at the indicated waiting times. (d) Horizon dependence at $\xi=kc\tau=0.1$, $t_w=5\tau$, comparing initialization at zero, restart, a two-state restart, and the zero predictor. (e) Two-state errors for the fitted, original learned, and physical-closure model banks at $t_w=5\tau$, $H=\tau$. (f) Achieved model-bank dimension minus the linear representation lower-bound dimension for arbitrary preparations, joint energy--flux observation, restart at $t_w=5\tau$, and tolerance $0.01$. Horizontal dotted lines in (a)--(e) mark $0.01$. All errors use the fixed initial-amplitude normalization. A low representation bound is necessary information about compressibility, not a guarantee that an autonomous closure of that dimension exists.}}
\label{fig6}
\end{figure*}

Figure~\ref{fig6}(a),(b) compares these achieved errors with the unresolved-response singular spectrum of Eq.~\eqref{eq:gram}. Waiting suppresses some observable directions, but the best linear representation bound can remain below the error attained by an autonomous model of the same dimension. Restricting preparations to equilibrium energy perturbations removes a substantial source of early-time ambiguity [Fig.~\ref{fig6}(c)], and averaging over a long window can dilute a transient error under the fixed RMS normalization [Fig.~\ref{fig6}(d)]. The measured error also changes with the declared two-state model bank [Fig.~\ref{fig6}(e)]. The positive dimension gap in Fig.~\ref{fig6}(f) therefore measures the difference between an achieved autonomous closure and a representation bound; it is not evidence for an untested globally minimal state count.

\subsection{Predictive branch combinations in a multichannel BTE}
\begin{figure*}[tp]
\centering
\includegraphics[width=\textwidth]{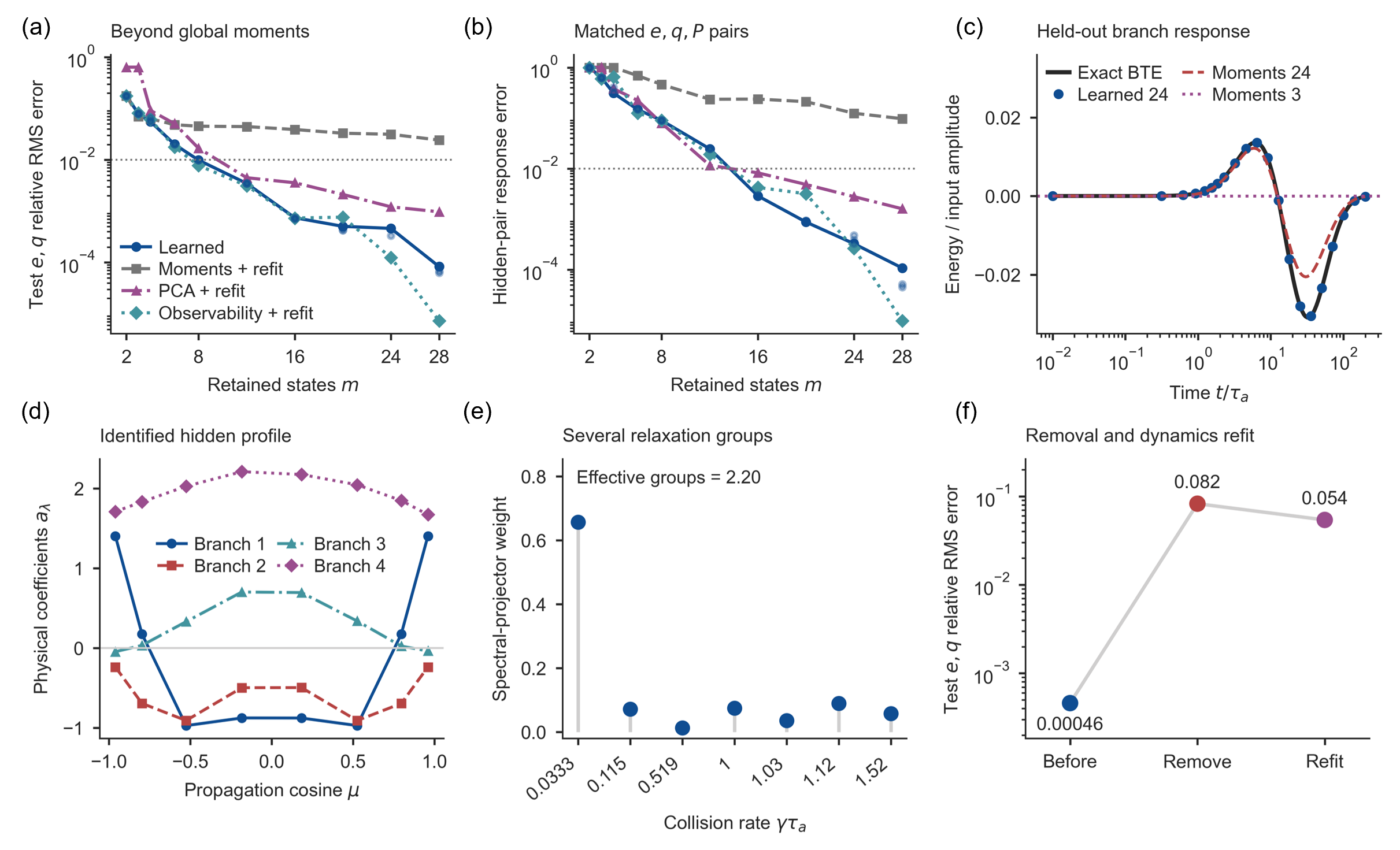}
\caption{\rev{Hidden predictive combinations beyond global moments in a multichannel BTE. A synthetic four-branch model contains 32 energy channels and reversible angular and interbranch scattering; $\tau_a$ is the angular relaxation time. (a) Joint energy--flux test error versus retained dimension for learned, fixed-moment, principal-component (PCA), and observability bases with refitted dynamics. (b) Difference-response errors for pairs matched in initial $e$, $q$, and $P$. Small learned-model symbols denote three optimization seeds; curves connect representative models. Dotted lines in (a),(b) mark $1\%$ error. (c) Energy response to a held-out branch preparation at $k=0.7$, compared with exact kinetics and rank-24 learned, rank-24 moment, and rank-three moment models. (d) Channel coefficients of a separately identified hidden direction across propagation cosine $\mu$ and branch index. (e) Its squared projector weights over positive collision-rate groups; the effective participation is $2.20$ groups. (f) Joint-observable relative RMS error before deletion, after deletion, and after refitting all retained dynamics: $4.61\times10^{-4}$, $8.24\times10^{-2}$, and $5.40\times10^{-2}$, respectively. This is a controlled synthetic kinetic model, not a material-specific phonon calculation.}}
\label{fig7}
\end{figure*}

The stronger test uses a synthetic energy BTE with four branches and eight angular channels per branch. The collision operator is reported to conserve energy and satisfy entropy-metric symmetry, while branch-dependent velocities couple several relaxation groups to the measured transport. Figure~\ref{fig7}(a),(b) compares learned, fixed-moment, PCA, and observability reductions with refitted dynamics. The learned reduction improves systematically with retained rank, whereas the tested global-moment hierarchy retains appreciable error. A held-out branch preparation produces a response missed by the three-moment model and captured by the rank-24 learned model [Fig.~\ref{fig7}(c)]. The observability baseline also performs strongly, so these results do not establish an exclusive advantage of a neural parametrization over linear response-based reduction.

The identified hidden direction is selected within the learned subspace subject to vanishing global $e$, $q$, and $P$. Its branch-dependent angular profile is therefore not one of those global moments [Fig.~\ref{fig7}(d)]. Its spectral-projector weights extend across several positive collision-rate groups, with an effective participation of $2.20$ groups [Fig.~\ref{fig7}(e)]; it is not a unique collision eigenmode. Deletion raises the joint-observable test error from $4.61\times10^{-4}$ to $8.24\times10^{-2}$, and refitting all retained dynamics reduces it only to $5.40\times10^{-2}$ [Fig.~\ref{fig7}(f)]. This identifies a predictive branch combination in the controlled synthetic model, without assigning it to a specific phonon material.

\subsection{Physical controls and state-consistency requirements}
\begin{figure*}[tp]
\centering
\includegraphics[width=\textwidth]{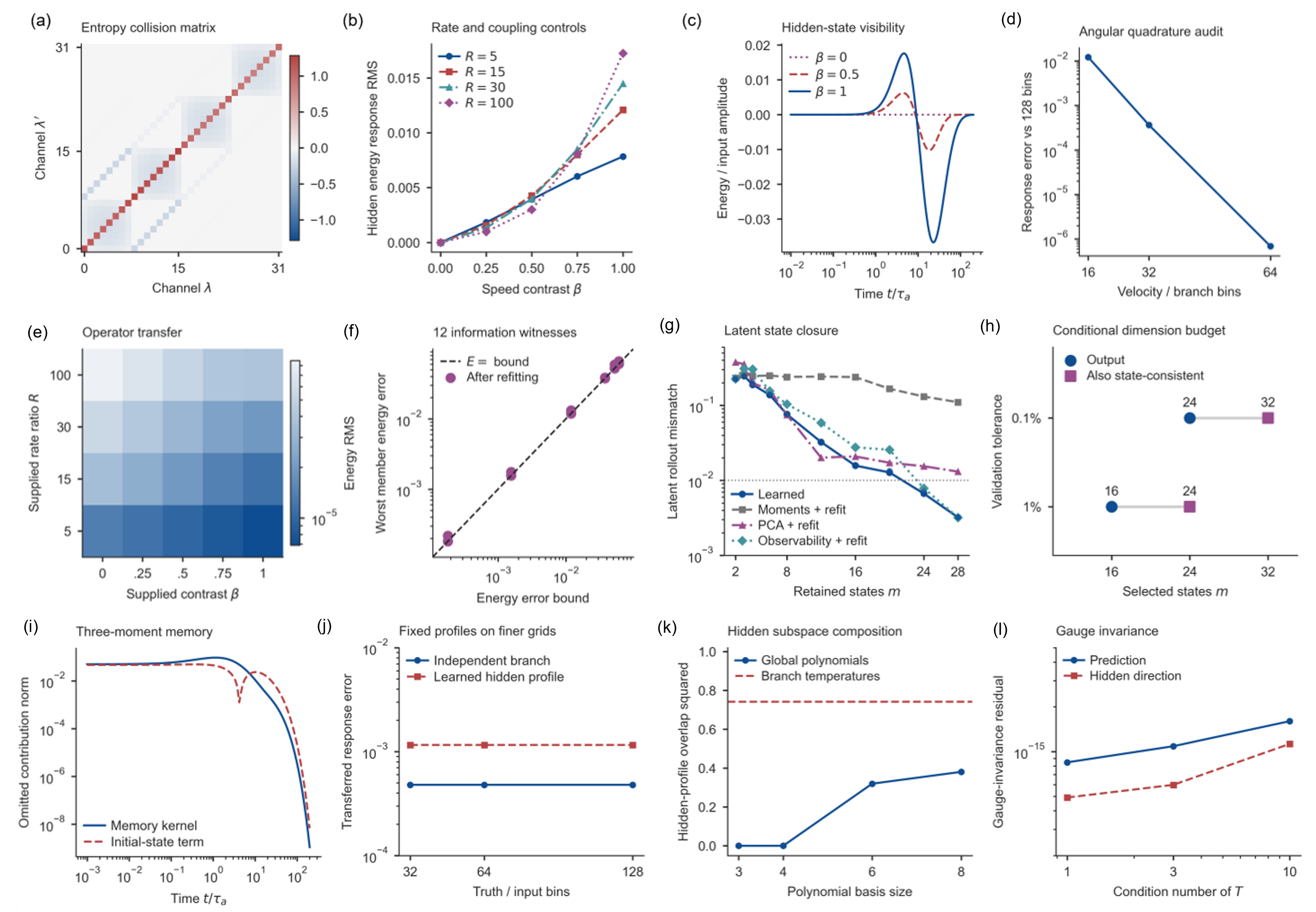}
\caption{\rev{Physical controls and autonomous-state tests for the multichannel BTE. (a) Signed symmetric collision representation in the entropy metric. (b) Hidden-response RMS under independent variation of the angular-to-slowest-branch rate ratio $R$ and branch-speed contrast $\beta$. (c) An initially unobserved branch combination becomes visible as $\beta$ increases. (d) Response convergence for 16, 32, and 64 channels relative to a 128-channel calculation. (e) Fixed-encoder transfer with the modified streaming and collision operators supplied. (f) Twelve new matched-input witnesses after hidden-direction deletion and dynamics refitting; the dashed line is the information lower bound. (g) Latent rollout mismatch for the four reduction methods; learned-model symbols include three seeds per rank. (h) Dimensions selected by observable accuracy alone and with an additional state-consistency test, at tolerances $1\%$ and $0.1\%$. (i) Norms of the exact omitted memory kernel and unresolved-initial-state term for a three-moment projection. (j) Transfer of fixed smooth learned profiles to finer angular quadratures. (k) Squared overlap of the identified direction with global polynomial subspaces of the indicated size and with the branch-temperature subspace. (l) Prediction and hidden-direction residuals under invertible coordinate transformations. All times use $\tau_a$. Parameter scans, quadrature tests, and matched-input witnesses are deterministic controls. Panel (e) tests transfer with known changed operators, not inference of unknown operators.}}
\label{fig8}
\end{figure*}

The controls in Fig.~\ref{fig8}(a)--(f) distinguish the presence of a hidden microscopic direction from its visibility in a chosen measurement. Changing the collision-rate ratio and branch-speed contrast independently changes its energy response; a direction that is unobserved when branch speeds coincide becomes visible as the contrast increases. Angular refinement checks the response against a finer reference. Transfer with a frozen encoder uses the changed operators as supplied inputs, and thus tests representation transfer rather than discovery of unknown dynamics. Twelve new matched-input witnesses after deletion and refitting obey the information lower bound.

In the tested model bank, observable accuracy alone selects a lower budget than the combined observable and state-consistency criteria. Figure~\ref{fig8}(g),(h) gives observable-only validation dimensions of 16 and 24 at tolerances $1\%$ and $0.1\%$, respectively; adding the latent re-encoding consistency test raises them to 24 and 32. The nonzero omitted memory and initial-state terms of a three-moment projection explain why an instantaneous low-order closure can fail [Fig.~\ref{fig8}(i)]. Fixed-profile transfer across angular quadratures [Fig.~\ref{fig8}(j)] and gauge invariance [Fig.~\ref{fig8}(l)] further constrain numerical and coordinate artifacts. The identified direction has only partial overlap with larger polynomial subspaces and about $0.74$ squared overlap with branch temperatures [Fig.~\ref{fig8}(k)], consistent with a mixed branch-and-angular kinetic state. These controls delimit the numerical finding; they do not establish a material-specific hidden phonon mode or validate the full nonlinear gated architecture.

\section{Scope and limitations of the inference}\label{sec:protocol}
A failed closure may reflect omitted states, an insufficient spatial stencil, inaccurate optimization, finite resolution, or incorrect boundary conditions. The analytic benchmark and linear baselines separate some of these causes; they do not eliminate all of them in a realistic material. A latent dimension is conditional on observations and preparations, not an intrinsic count of material degrees of freedom. There is no necessary relation between network depth and thermodynamic scale, and arbitrary spatial averaging does not establish a renormalization-group transformation.
The numerical results establish a map of predictively sufficient models within a stated bank and identify hidden directions whose removal produces certified errors. The three-velocity system validates recovery of a known higher-moment direction; the multichannel model shows how a predictive state can instead combine several branches and collision-rate groups. Establishing an additional material-specific transport channel requires independently validated microscopic scattering and observable responses. That physical identification, as well as nonlinear scale-conditioned state discovery, remains beyond the present controlled benchmarks.

\section{Conclusion}
We have investigated the emergence of effective transport dynamics from the perspective of predictive state sufficiency. Rather than assuming that microscopic kinetic theory reduces through a predetermined hierarchy of macroscopic constitutive equations, we ask which microscopic distinctions must remain represented for a selected set of observables to evolve autonomously. This viewpoint turns coarse-graining into a testable dynamical problem: a proposed reduction is incomplete whenever distinct microscopic preparations, indistinguishable in the retained state, produce different observable futures.
The theoretical framework establishes two complementary measures of missing predictive information. Conditional future variance quantifies the irreducible prediction error associated with a given state representation, while finite-horizon observability identifies unresolved microscopic directions capable of influencing subsequent measurements. Together, these concepts distinguish dynamical relevance from instantaneous reconstructability or relaxation time alone. They also clarify an essential limitation of low-dimensional closure: eliminating microscopic variables generally produces memory and dependence on unresolved initial conditions. Consequently, a reduced model cannot be considered autonomous solely because it accurately reproduces a restricted collection of observable trajectories.
We demonstrated these principles using controlled kinetic benchmarks. In the exactly solvable three-velocity system, two preparations with identical initial energy and heat-flux fields exhibit different subsequent temperature evolution because of an unresolved higher-order moment. The learned full-rank representation recovers the corresponding physical subspace, whereas removing this direction produces prediction errors that cannot be eliminated by refitting the remaining dynamics. Independent preparation pairs, finite-pair information bounds, and coordinate-invariant subspace tests establish that this failure originates from missing predictive information rather than an unfavorable choice of latent coordinates. The resulting model-bank dimension maps show that the states required for accurate prediction depend on the observation window, initialization protocol, physical scale, and prescribed tolerance.
A more complex synthetic Boltzmann system containing 32 kinetic channels reveals that predictive states need not correspond to individual conventional moments or isolated collision eigenmodes. The identified hidden direction combines branch-dependent angular structure with contributions from multiple relaxation groups. Removing this direction increases the joint-observable relative error from \(4.61\times10^{-4}\) to \(8.24\times10^{-2}\), while refitting the remaining dynamics reduces the error only to \(5.40\times10^{-2}\). These results provide a controlled example of a physically consequential kinetic subspace that is not represented by the lowest global moments. Furthermore, imposing autonomous-state consistency raises the validated state budget beyond that selected from observable accuracy alone, demonstrating that accurate output prediction does not necessarily imply a dynamically closed internal representation.
The resulting physical interpretation differs from a universal progression through Fourier, Cattaneo–Vernotte, and higher-order transport theories. These descriptions are instead candidate reductions whose validity depends on the accessible microscopic preparations, spatiotemporal observations, and predictive accuracy. The effective state dimension is therefore an operational measure of dynamical complexity, not an intrinsic number of degrees of freedom belonging to a material. A decrease in this dimension may reflect physical relaxation and scale separation, but it can also arise from restricted initialization, delayed observation, or reduced measurement sensitivity. Distinguishing these mechanisms is essential when interpreting learned transport laws.
The present numerical evidence concerns linear near-equilibrium kinetic systems with explicitly controlled collision operators. Although the formulation accommodates nonlinear encoders, scale-conditioned state selection, and memory-augmented evolution, their predictive sufficiency has not yet been established for nonlinear or material-specific phonon transport. Such extensions will require independent microscopic scattering information, physically admissible preparation ensembles, and validation against observable responses.
More broadly, our results suggest that the emergence of macroscopic transport laws can be characterized by the progressive loss of \emph{predictively relevant} microscopic information. The central question is not simply how many microscopic variables can be discarded, but whether the distinctions that remain unresolved can still alter the future. By linking this question to measurable response differences, information bounds, and physically constrained state identification, the framework provides a route toward constructing effective transport theories whose degrees of freedom are determined by dynamical necessity rather than prescribed in advance.

\section*{Acknowledgments}
This work was supported by the Adolf Martens Fellowship (Grant n. BAM-AMF-2025-1).

\appendix
\section{Exact autonomous closure and a consistency error bound}\label{app:closure}
For a fixed Fourier wave number (or a fixed finite spatial discretization), write $\dot x=Lx$ and let $E$ be a full-row-rank linear encoder. An exact autonomous linear state $z=Ex$ for all initial states exists if and only if
\begin{equation}\label{eq:intertwine}
 EL=L_rE,
 \qquad\text{equivalently}\qquad L(\ker E)\subseteq\ker E.
\end{equation}
Necessity follows by differentiating $Ee^{tL}=e^{tL_r}E$ at $t=0$. Conversely, invariance of $\ker E$ makes $L_r(Ex)=ELx$ independent of the representative $x$, and exponentiation gives the intertwining identity. Exact prediction of an instantaneous observable $B_ox$ additionally requires $B_o=D_oE$. For a local encoder shared by a transport PDE, these requirements must hold for every admitted $k$, not only one fitted mode. For $L(k)=C_B-ikV$, intertwining for a continuum of $k$ requires compatible reductions of both $C_B$ and $V$.

This condition concerns a quotient state, rather than merely an invariant reconstruction subspace. In entropy-whitened coordinates with symmetric streaming and collision matrices, a common invariant subspace has an invariant orthogonal complement; without that structure these notions must not be interchanged.

For $E=U^\dagger G$, $U^\dagger GU=I$, $P=UE$, and a fitted $L_r$, define
\begin{equation}
 R=EL-L_rE.
\end{equation}
The encoded microscopic state $z_*(t)=Ex(t)$ and reduced rollout $\widehat z(t)=e^{tL_r}Ex(0)$ satisfy
\begin{equation}\label{eq:defect-duhamel}
 z_*(t)-\widehat z(t)=\int_0^t e^{(t-s)L_r}R e^{sL}x(0)\,ds.
\end{equation}
If both semigroups are contractive in their declared metrics, then
\begin{equation}\label{eq:defectbound}
 \|z_*(t)-\widehat z(t)\|_2
 \leq t\,\|RG^{-1/2}\|_2\,\|x(0)\|_G.
\end{equation}
For Galerkin dynamics $L_r=ELU$, $R=EL(I-P)$. The same coupling that drives unresolved-state feedback therefore controls failure of autonomous re-encoding consistency. Equation~\eqref{eq:defectbound} is a sufficient estimate, not a sharp lower bound. Agreement at a few sampled times cannot prove $R=0$.

For the physical decoder $D_o=B_oU$, the observable discrepancy splits exactly into
\begin{equation}
 B_ox(t)-B_oU\widehat z(t)
 =B_o(I-P)x(t)+B_oU[z_*(t)-\widehat z(t)].
\end{equation}
Small measured output error need not imply that either contribution or the latent defect is small. An instantaneous output may even be insensitive to a large latent discrepancy. For nonlinear encoders the corresponding exact condition is $D\mathcal E(x)F(x)=F_r(\mathcal E(x))$ on the admissible state set, with the right-hand side constant on every encoder fiber. No nonlinear closure theorem is inferred from the linear tests.

\section{Admissible information bounds and observation protocols}\label{app:bounds}
The finite-pair and conditional-variance identities require a fixed positive-semidefinite output weight $W$ and finite second moments. Strict positive definiteness is needed only when the weighted quantity is to be a norm on the full output space. For complex Fourier data the covariance and inner products use conjugate transpose; alternatively, stack real and imaginary parts.

The singular-value bound applies to a full unresolved perturbation ball. With $BN=0$ and $N^\dagger GN=I$, suppose every state $X_0+aN\xi$, $\|\xi\|_2\leq1$, is admissible. For positive population variables with $X_{0i}>0$, a sufficient condition is
\begin{equation}
 |a|\,\|N_{i,:}\|_2<X_{0i}\quad\text{for every }i.
\end{equation}
Physical real Fourier perturbations must include the conjugate mode. The bound must be applied to that realified admissible space or to a consistently normalized complex-amplitude convention.

A linear encoding of at most $r$ additional unresolved coordinates followed by a linear future decoder has rank at most $r$ on this ball. The optimal worst-case response error is $|a|\sigma_{r+1}(\mathsf O_HN)$. If the preparation set is an ellipsoid, its amplitude map must be included before taking singular values. If only finitely many preparations or a smaller positivity-truncated set are tested, the full-ball bound cannot be imposed without establishing that the required witness directions belong to the set. An encoder retaining $b=\operatorname{rank}B$ independent fields has total budget at least $b+r$ in this restricted linear comparison; $r$ does not count those retained fields again. This is an information bound, not an existence result for a local autonomous PDE.

For a continuous-window RMS criterion with a fixed output scaling $W_o$, the relevant Gramian is
\begin{equation}\label{eq:rmsgram}
 Q_{w,H}=\frac1H\int_{t_w}^{t_w+H}
 e^{tL^\dagger}B_o^\dagger W_o B_o e^{tL}\,dt.
\end{equation}
The squared singular values on the normalized unresolved subspace are the eigenvalues of $N^\dagger Q_{w,H}N$. Discrete quadrature must include the matching time weights. A supremum-in-time tolerance, this RMS tolerance, and a trajectory-relative RMS denominator define different criteria. A worst-case preparation-ball error uses an operator norm; an ensemble-average error depends on the preparation covariance and is generally a trace quantity. These criteria cannot share one lower-bound curve without a normalization check.

If $x(0)=S_0\xi$, a restart at $t_w$ receives $Ee^{t_wL}S_0\xi$, while a model initialized at zero receives $ES_0\xi$. Retaining the same original ball induces an ellipsoid at the restart time. Renormalizing that ellipsoid to a new unit ball changes the preparation protocol. Re-encoding at $t_w$ also supplies kinetic information that the earlier reduced rollout may have lost. Comparisons in Figs.~\ref{fig5} and \ref{fig6} must preserve these distinctions.

\section{Three-velocity dispersion, elimination, and the Fourier baseline}\label{app:diffusion}
Writing $r=P-c_s^2e$ and $d=c^2-c_s^2>0$, the exact generator in $(e,q,r)$ is
\begin{equation}
 L_{eqr}(k)=\begin{pmatrix}
 0&-ik&0\\-ikc_s^2&-\tau^{-1}&-ik\\0&-ikd&-\tau^{-1}
 \end{pmatrix}.
\end{equation}
Its characteristic polynomial is
\begin{equation}\label{eq:cubic-review}
 s(s+\tau^{-1})^2+c^2k^2s+\tau^{-1}c_s^2k^2=0.
\end{equation}
The root emerging from the conserved mode satisfies
\begin{equation}
 s_{\mathrm{hyd}}(k)=-\tau c_s^2k^2+O(k^4).
\end{equation}
Thus the diffusivity is $\alpha=\tau c_s^2$. This is a long-wavelength, long-time reduction after kinetic transients; it does not reproduce arbitrary initial flux and hidden-moment responses. At $k=0$, both non-energy modes relax at $\tau^{-1}$, so there is no intrinsic separation of their collision lifetimes that would by itself justify dropping only $r$.

For weights $w_+=w_-=1/6$, $w_0=2/3$, which yield the reported couplings, use $\delta e=e-e_0$ and the quadratic metric $G_{ii}=1/(w_i e_0)$. Define
\begin{equation}
 z_e=\frac{\delta e}{\sqrt{e_0}},\quad
 z_q=\frac{q}{c_s\sqrt{e_0}},\quad
 z_r=\frac{r}{c_s\sqrt d\sqrt{e_0}}.
\end{equation}
Then $A_{eq}=c_s=c/\sqrt3$, $A_{qr}=\sqrt d=c\sqrt{2/3}$, and $D=\operatorname{diag}(0,\tau^{-1},\tau^{-1})$. These values follow from the specified weights, not from symmetry alone. The saved simulation configuration confirms these preparation weights; the numerical conventions are recorded in Appendix~\ref{app:reporting}.

Projecting directly onto $z_e$ sets both transport and collision matrices to zero. In contrast, quasistatic elimination of the fast normalized variables gives $z_q\simeq-\tau c_s\partial_xz_e$ at leading gradient order and hence $\partial_tz_e\simeq\tau c_s^2\partial_x^2z_e$. The effective diffusion term is generated by eliminated feedback. It is absent from a one-state first-order Galerkin generator. A fitted Fourier model in the bank is therefore a separate asymptotic model class with a gradient-dependent flux decoder.

\section{Entropy, realizability, and coordinate-covariant audits}\label{app:physical}
For the three-velocity energy-population benchmark, a compatible convex entropy deficit is
\begin{equation}
 \mathcal H[f]=\int\sum_i f_i\ln\frac{f_i}{w_i e_0}\,dx.
\end{equation}
At positive populations, periodic streaming integrates to zero and the collision term obeys
\begin{equation}
 \left.\frac{d\mathcal H}{dt}\right|_{\mathrm{coll}}
 =-\frac1\tau\int\sum_i(f_i-w_ie)
 \ln\frac{f_i}{w_ie}\,dx\leq0.
\end{equation}
The last inequality holds termwise because $(a-b)\ln(a/b)\geq0$ for $a,b>0$. Conservation removes the common logarithmic and linear terms. The quadratic Hessian at $f_i^0=w_i e_0$ is $G_{ii}=1/(w_i e_0)$. This classical population entropy is distinct from the Bose occupation entropy in the main text. A positive diagonal metric on synthetic channels must not be identified with a material-specific phonon entropy without the requisite occupation and quadrature definitions.

For a population collision generator $C_B$, energy conservation and $G$-self-adjoint negative semidefiniteness establish a quadratic contraction, but do not alone establish positivity. In the convention $\dot f=C_Bf$, positivity of the full linear population dynamics requires nonnegative off-diagonal entries (the Metzler condition). Together with $\boldsymbol1^TC_B=0$, a positive equilibrium $C_Bw=0$, and $G=\operatorname{diag}(1/w_i)$, detailed balance can be expressed as $(C_B)_{ij}w_j=(C_B)_{ji}w_i$. Signed entries of the whitened matrix in Fig.~\ref{fig8}(a) must be interpreted in their stated basis: a matrix shown after a general basis rotation is not tested for population positivity by its entry signs. A general linearized phonon operator also need not be a population Markov generator; in that case the justified claim is near-equilibrium linear response, not positivity for arbitrary amplitudes.

A Galerkin decoder $f^0+Uz$ need not remain nonnegative even if the microscopic model does. The quadratic entropy proof therefore establishes stability in a metric, not global realizability or a nonlinear $H$ theorem. Energy normalization, positivity of all prepared microscopic states, and positivity of decoded states are separate tests. A centered conservative spatial derivative is skew-adjoint on a periodic grid, while a numerical time integrator must independently preserve the required stability; the continuous-time identity is not a guarantee for an arbitrary time step.

Finally let $z'=Tz$ for invertible $T$. Then
\begin{equation}
 A'=TAT^{-1},\quad D'=TDT^{-1},\quad
 M'=T^{-\dagger}T^{-1}.
\end{equation}
Physical symmetry and dissipation transform as
\begin{equation}
 M'A'=A'^\dagger M',\qquad
 M'D'=D'^\dagger M'\succeq0.
\end{equation}
For a more general dissipative generator $L_r'$, the condition is $L_r'^\dagger M'+M'L_r'\preceq0$. Euclidean symmetry of $A'$ or $D'$ is not invariant under a nonorthogonal change of coordinates. The decoder transforms as $U'=UT^{-1}$, and norms, subspace comparisons, and conditioning corrections must use the transformed metric. Orthogonal transformations are the special case $M'=I$.

\section{Numerical implementation and verification}\label{app:reporting}
This appendix records the completed linear benchmarks underlying Figs.~\ref{fig2}--\ref{fig8}. The implementation, raw trajectory arrays, frozen model coefficients, split metadata, and optimization histories are retained in the local reproducibility files. The general nonlinear and gated designs in Appendixes~\ref{sec:architecture} and \ref{app:protocol} are separate prospective extensions. The verification below re-evaluates frozen parameters and selected direct kinetic calculations; it does not constitute a new training campaign from random initialization.

\subsection{Three-velocity reference and preparation ensemble}
The velocity order is $(+c,0,-c)$, with $c=\tau=e_0=1$ and weights $(1/6,2/3,1/6)$, so that $c_s^2=1/3$ and $G=\operatorname{diag}(6,3/2,6)$. For Fig.~\ref{fig2}(a)--(f), the periodic domain has $L=2\pi$, $k=1$, and $N_x=128$. The two initial distributions are $we_0\mp(a/2)\cos(kx)(1,-2,1)^T$, with $a=10^{-3}$. Fourier matrix exponentials provide the reference. A separate pseudospectral kinetic RK4 integration uses maximum step $0.005\tau$, shortened at requested output times, without imposing a moment closure. The early-time exponent is fitted on 41 observations in $10^{-4}\leq t/\tau\leq10^{-3}$. Its recorded value, $1.99975835$, is consistent with an independent matrix-exponential calculation, $1.99975835$. The RK4 cosine separation at $t=\tau$ is $-4.889014580847309\times10^{-4}$; the exact propagator gives $-4.889014580738867\times10^{-4}$. Their small discretization difference is retained rather than treated as a disagreement of identical reference calculations.

The training ensemble contains 2000 trajectories generated with pseudorandom seed 20261012. The five preparation families---energy, flux, higher moment, mixed Fourier modes, and hidden pairs---contain 400 trajectories each. The 1400/300/300 training, validation, and test split is assigned by complete preparation before fitting. The 400 hidden-pair members form 200 pairs, distributed as 140/30/30 pairs across those splits. Both members always remain together. The test set contains 150 in-distribution and 150 held-out-band trajectories. The reserved physical wave-number intervals are $[0.3,0.6]$ and $[3,5]$. Training and validation preparations exclude those intervals for every occupied harmonic.

Each realization has period $L=2\pi/k_0$, $N_x=128$, and at most four harmonics, with every occupied $mk_0$ in $[10^{-2},10]$. Candidate base wave numbers are 65 logarithmic points in that interval, supplemented by $0.1$ and $1$. Reference perturbations are represented by complex amplitudes $b_m$ according to
\begin{equation}
 \delta f(x,t)=\operatorname{Re}\sum_m b_m(t)e^{imk_0x}.
\end{equation}
Thus a normalized FFT coefficient at $+mk_0$ is $b_m/2$; spatial RMS norms carry the matching Parseval factor $1/2$. Random preparations are rescaled to a logarithmically sampled bound between $10^{-3}$ and $5\times10^{-3}$ on the relative population perturbation. The explicitly included benchmark pair has maximum initial relative perturbation $0.003$. Positivity is checked on the microscopic populations, separately from reduced-model entropy contraction. Fifty saved preparations also have $N_x=256$ refinement data; these band-limited grid tests do not establish convergence for arbitrary angular or spatial fine structure.

There are 301 exact observations per trajectory. In units of $\tau$, their grid is the sorted union of 201 points from 0 to 2, 20 logarithmic points from $10^{-4}$ up to but excluding $10^{-2}$, 79 geometric points from 2.01 to $T$, and the point 5, where $T=\max(8,15/k_0^2)$. The array dimensions are $2000\times128\times3\times301$; complex modal amplitudes are stored separately. The equilibrium is the known center. Initial RMS scales are fitted on training members only. The saved energy and flux scales used in trajectory fitting are $s_e=6.831908008510406\times10^{-4}$ and $s_q=2.588542838037134\times10^{-4}$; these remain fixed during validation and testing.

\subsection{Three-velocity fitting, deletion, and certificates}
Ranks $m=1,2,3$ are fitted separately with seeds $17,29,41,53,67$. In entropy-whitened coordinates, the first basis vector is $\sqrt w$. A fixed geometric frame for its orthogonal complement, generated with seed 91027, does not assign the free channels to flux or a higher moment. The energy transport row equals the decoded microscopic flux; the remaining transport block is symmetric and the non-energy dissipation block is a Cholesky product. The anchored first-order $m=1$ model has no free parameters and is static. It is not the fitted Fourier baseline used below.

Trajectory errors use the fixed training scales $(s_e,s_q)$, exact modal Parseval averaging, and normalized observation weights: one half of the weight is assigned to observations with $t\leq2\tau$ and one half to later observations. The hidden-pair difference-response loss has coefficient one and uses its recorded pair amplitude. Fits use 28 selected observations, while validation and test scoring use all 301. Training-only finite-difference regression initializes the unknown matrices. L-BFGS-B uses analytic matrix-exponential Fr\'echet derivatives, at most 250 iterations, maximum 40 line-search steps, gradient tolerance $10^{-12}$, and function tolerance $10^{-20}$. The free transport entries lie in $[-2,2]$, log Cholesky diagonals in $[-3,2]$, and the rank-three off-diagonal Cholesky parameter in $[-2,2]$. The best validation checkpoint is retained; the two-state fits additionally compare three initializations per seed. These settings describe a constrained linear fit, not a deep network or soft gate.

The hidden line for Figs.~\ref{fig3} and \ref{fig4} is identified from the leading singular vector of normalized \emph{training} pair encodings and then lifted to the whitened physical space. One two-state control holds the retained basis fixed and refits its remaining dynamics. A second control refits the remaining encoder direction as well as the dynamics, with three starts per seed and the same validation-only checkpoint rule. A new witness is required after that encoder changes; the original $(e,q)$-matched pair need not remain latent-matched.

The 225 certificates comprise 30 original test pairs per seed and 15 new pairs per seed matched to the refitted free encoder. The latter use $k=0.01,0.1,1,3,10$ and three backgrounds, with pair-difference population bound $0.002$ and background bound $0.001$. They are evaluated at 161 points on $[0,4\tau]$ with normalized trapezoidal time weights. Original-pair certificates instead retain their saved 301-point, half-transient/half-late-time weights. Each certificate uses exactly the same fixed output scales, spatial Parseval factor, and temporal weights for its separation $d$ and prediction error.

For nonzero encoded mismatch $\epsilon_z$, the recorded lower bound is $\max(0,d-L_h\epsilon_z)/2$. Here $L_h$ is the largest singular value of the time-weighted reduced response map, maximized over occupied harmonics. The nominal $d/2$, corrected bound, actual error of each member, mismatch, and admissible amplitude are all retained. Re-evaluation of all 225 certificates reproduces the saved values and respects both energy-only and joint-observable corrected bounds. The general invertible-coordinate tests use $M'=T^{-T}T^{-1}$ and the covariant symmetry conditions in Appendix~\ref{app:physical}. The maximum replayed prediction change is $4.15\times10^{-15}$; the transport and dissipation metric-symmetry residuals are below $4.1\times10^{-16}$.

\subsection{Scale maps and matched information bounds}
The grids contain 61 logarithmic $\xi=kc\tau$ values in $[10^{-2},10]$, 71 logarithmic horizons $H/\tau$ in $[10^{-2},10^5]$, waiting times $0,1,5$, and tolerances $0.1,0.03,0.01,0.003,0.001,10^{-4}$. Both arbitrary and equilibrium-energy preparations and both energy-only and joint energy--flux observations are evaluated.

The main bank contains a one-field Fourier model, a separately fitted two-state model at each $\xi$, and a frozen learned three-state model. The Fourier diffusivity $\alpha=0.3329942100282546$ is fitted from training data, with algebraic flux $q=-\alpha\partial_xe$. The two-state fits use three deterministic initializations at each wave number (183 runs in total), L-BFGS-B with maximum 160 iterations, maximum 80 line-search steps, gradient tolerance $10^{-9}$, and function tolerance $10^{-16}$. Their fitting time grid extends to $\max(11,3/\xi^2)$ and assigns equal total weight before and after $2\tau$. Once selected on validation data, a generator is frozen across all evaluated $H$, $t_w$, and $\varepsilon$. These are wave-number-conditioned fits, not one globally shared local PDE generator; no fit is performed separately for each requested forecast horizon.

The map norm is continuous-window RMS, not the snapshot-weighted training loss. For a single Fourier amplitude $\delta f=10^{-3}\sqrt W\,y$, the initial ball is $\|y\|_2\leq1$ and the physical spatial amplitude scale is $a_{\mathrm{RMS}}=10^{-3}/\sqrt2$. Output scales are $a_{\mathrm{RMS}}$ for energy and $c_sa_{\mathrm{RMS}}$ for flux. With the corresponding normalized observation matrix $C$, the error Gramian is
\begin{equation}\label{eq:E-map-gram}
 Q_R=H^{-1}\int_0^H R(t)^\dagger R(t)\,dt,
 \qquad E_{\mathrm{ball}}=\sqrt{\lambda_{\max}(Q_R)}.
\end{equation}
All input amplitudes in this expression refer to the original time-zero ball. Restart uses encoder $B^Te^{t_wL}$, rather than renormalizing the decayed restart state. The energy row whose nullspace enters the representation bound is likewise $b_e^Te^{t_wL}$ for restart, and $b_e^T$ for initialization at zero. The truth map, admissible original ball, restart map, normalized outputs, and continuous-window weights therefore agree between the lower-bound and achieved-error calculations.

Exponential integrals evaluate the Gramians, with small-argument series to avoid cancellation. The nearly exact full-rank error is certified by a contractive Duhamel bound instead of subtracting nearly equal Gramians. Replaying all saved main-bank error entries and dimension cells gives identical arrays. Four direct matrix-exponential quadrature checks, including $H/\tau=10^5$, differ from the analytic Gramians by less than $3.7\times10^{-16}$. Cells where the zero predictor also passes remain explicitly flagged. None of these calculations converts a linear representation lower bound into an existence theorem for a local autonomous closure.

\subsection{Explicit multichannel operator}
The multichannel model has four branches with equilibrium fractions
\begin{equation}
 h=(0.38,0.27,0.22,0.13),\qquad
 c=(1,0.72,0.43,0.22).
\end{equation}
For $n_a$ Gauss--Legendre nodes $\mu_\alpha$, let $\omega_\alpha$ be their weights divided by two, so $\sum_\alpha\omega_\alpha=1$. The equilibrium population and velocity of channel $(b,\alpha)$ are
\begin{equation}
 w_{b\alpha}=h_b\omega_\alpha,\quad
 v_{b\alpha}=[\bar c+\beta(c_b-\bar c)]\mu_\alpha,
 \quad\bar c=\sum_bh_bc_b.
\end{equation}
The main calculation uses $n_a=8$ (32 channels), $\beta=1$, and $R=30$. Time is measured in the angular relaxation time $\tau_a$, and wave number in $(c_1\tau_a)^{-1}$, with $c_1=1$. The symmetric interbranch conductance matrix is
\begin{equation}
 K=\begin{pmatrix}
 0&1&0.08&0.02\\
 1&0&0.30&0.04\\
 0.08&0.30&0&0.10\\
 0.02&0.04&0.10&0
 \end{pmatrix}.
\end{equation}
Set $(Q_0)_{bb'}=h_bK_{bb'}$ off the diagonal and $(Q_0)_{bb}=-\sum_{b'\ne b}(Q_0)_{b'b}$. For $H_b=\operatorname{diag}h$, define $D_b^0=-H_b^{-1/2}Q_0H_b^{1/2}$. If $\gamma_{\min}^0$ is its smallest positive eigenvalue, put $D_b=D_b^0/(R\gamma_{\min}^0)$. The whitened dissipation is
\begin{equation}\label{eq:E-collision}
 D=D_b\otimes I_{n_a}
 +I_4\otimes(I_{n_a}-ss^T),\qquad s_\alpha=\sqrt{\omega_\alpha}.
\end{equation}
With $W=\operatorname{diag}w$, the population collision operator is $C_B=-W^{1/2}DW^{-1/2}$ and the entropy metric is $G=W^{-1}$. This operator has nonnegative off-diagonal population rates, conserved energy, positive equilibrium, and detailed balance; it is a positivity-preserving synthetic population generator, not merely a stable arbitrary linear matrix. Reconstructed machine-readable operators agree with the saved matrices. The zero-centered signed matrix displayed in Fig.~\ref{fig8}(a) is $D$, not $C_B$. No material-specific Bose collision operator is inferred from this construction.

The controls vary $R=5,15,30,100$ independently of $\beta=0,0.25,0.5,0.75,1$. Total channel counts 16, 32, 64, and 128 mean $n_a=4,8,16,32$, respectively. Refinement retains the same branch conductance and continuous angular relaxation rule and changes only the angular quadrature. Convergence and fixed-profile transfer are tested for the stated smooth branch/angular preparations; independently drawn high-order angular randomness is not the same preparation across grids.

\subsection{Multichannel data, training, and model order}
Seed 20261015 generates 2000 single-mode preparations, with counts 200, 200, 200, 400, 400, 400, and 200 for equilibrium energy, global flux, global second moment, branch temperatures, branch fluxes, full angular perturbations, and $(e,q,P)$-matched branch-pair members. The split is 1400/300/300, containing 70/15/15 complete hidden pairs. Training wave numbers are $0.03,0.1,0.3,1,3$; additional test wave numbers are $0.02,0.06,0.2,0.7,2$. Physical populations are $w+10^{-4}\sqrt W\operatorname{Re}(y_0e^{ikx})$. Their recorded maximum initial relative perturbation is $5.53\times10^{-4}$.

The 122 observation times are $0$, 40 equally spaced points from $0.01$ to $2$, and 81 geometric points from $2.05$ to $200$, in units of $\tau_a$. Sixty-one microscopic modal snapshots are also saved. Time weights assign fractions $0.4$, $0.3$, and $0.3$ to $t\leq2$, $2<t\leq20$, and $t>20$, respectively, uniformly over the sampled times in each interval. These define a discrete weighted \emph{modal-amplitude} norm. They are not the continuous-window norm of Eq.~\eqref{eq:E-map-gram}; the same fixed discrete norm is used on both sides of every multichannel certificate.

Microscopic operator regression uses training initial states and their exact microscopic time-derivative labels. A predictive response Gramian is inferred from training observable responses and initializes the learned and observability bases. Thus initialization uses derivative supervision and known physical observables; it is not identification from temperature snapshots alone. The PCA and moment baselines use training covariance and global velocity polynomials, respectively. All bases anchor the conserved energy direction. Fitted dynamics enforce symmetric transport, Cholesky-positive non-energy dissipation, and the decoded energy-flux row.

Ranks are $2,3,4,6,8,12,16,20,24,28$, with a full-rank 32-state operator-regression reference. The full-rank reference is not a learned reduced model. Learned bases use seeds $17,29,41$; each fixed-basis baseline uses seed 17. The training covariance is the sum of trajectory and pair-difference covariances, with pair coefficient one. The objective adds a latent re-encoding residual with coefficient $0.02$. Each wave-number group has equal total weight. Fits use 39 selected time points; validation and test metrics use all 122.

The implemented optimizer is an Adam-type retracted adaptive-gradient update, not the three-velocity L-BFGS-B optimizer. Its first and second moment coefficients are $0.9$ and $0.999$, denominator offset is $10^{-9}$, and initial step sizes for the basis, free transport block, and Cholesky parameters are $0.008,0.004,0.004$. The steps are multiplied by 1, 0.25, and 0.06 over the first 50\%, next 30\%, and final 20\% of iterations. The encoder is retracted by polar orthonormalization with its energy column fixed. Free transport entries, off-diagonal Cholesky parameters, and log diagonal Cholesky parameters are clipped to $[-2,2]$, $[-4,4]$, and $[-6,2]$, respectively. Learned bases use 240 updates and fixed-basis baselines 180. Validation checkpoints every ten updates select the exported model. No test error selects a checkpoint or rank.

Let $F_k(t)=e^{tL_k}$ and $M_k=-ikA-D_r$. For covariance $\Sigma_k=\Sigma_k^{\mathrm{traj}}+\Sigma_k^{\mathrm{pair}}$, the state-consistency metric is
\begin{equation}\label{eq:E-state-metric}
 \eta_z^2=
 \frac{\sum_{k,j}\omega_j\operatorname{tr}(R_{kj}\Sigma_kR_{kj}^\dagger)}
 {\sum_{k,j}\omega_j\operatorname{tr}(Z_{kj}\Sigma_kZ_{kj}^\dagger)},
\end{equation}
where $R_{kj}=e^{t_jM_k}B^T-B^TF_k(t_j)$ and $Z_{kj}=B^TF_k(t_j)$. Observable and pair-response metrics have their respective trajectory and pair covariances and true-response denominators. Observable selection requires both relative errors below $\varepsilon_o$; state-consistent selection additionally requires $\eta_z\leq\varepsilon_z$. Figure~\ref{fig8}(h) uses $\varepsilon_z=\varepsilon_o$ at the two numerical tolerances $0.01$ and $0.001$, without treating the criteria as identical observables. Recomputed dimensions are 16 and 24 under the observable tests, and 24 and 32 with state consistency. They are conditional on this rank bank and normalization.

\subsection{Hidden direction, deletion, and verification scope}
The parent used for physical identification has rank 24. The identified unit vector $h_{\mathrm{hid}}$ maximizes the training energy-response quadratic form within $\operatorname{range}B\cap\ker B_{eqP}$, where $B_{eqP}$ collects the microscopic global energy, flux, and second-moment rows. The direction is reidentified from the training response arrays, rather than assigned to a raw latent index. For distinct collision-rate groups $\gamma$, define
\begin{equation}
 p_\gamma=\|\Pi_\gamma h_{\mathrm{hid}}\|_2^2,
 \qquad N_{\mathrm{grp}}=\left(\sum_\gamma p_\gamma^2\right)^{-1}.
\end{equation}
Eigenvalues separated by at most $10^{-8}$ are grouped before forming their spectral projector $\Pi_\gamma$. The energy-group weight vanishes up to roundoff. The result is $N_{\mathrm{grp}}=2.195254062$, reported as 2.20; the squared overlap with branch temperatures is $0.7423169712$.

Deleting this direction reduces the parent from 24 to 23 states. The retained encoder is \emph{fixed} in this multichannel refit; all remaining dynamical parameters are reoptimized for 200 updates using three starts with seeds $17,29,41$, selected by validation loss. This tests the retained subspace only; it does not establish a lower bound for every 23-state encoder. Re-evaluating the frozen test scores gives relative joint-observable RMS errors $0.000460652089$, $0.0823537712$, and $0.0539938577$ before deletion, after deletion, and after refitting, respectively.

Twelve new matched-input witnesses use six wave numbers $0.02,0.06,0.2,0.7,2,3$ and two backgrounds. Their mismatch is measured in the whitened latent norm. Contractivity gives the response Lipschitz bound $L_h\leq\|C_oB\|_2$ for the normalized discrete time weights; this corrects $d/2$ to $\max(0,d-L_h\epsilon_z)/2$. Fresh matrix-exponential truth and frozen-model responses satisfy every corrected bound. Multichannel coordinate tests also use $M'=T^{-T}T^{-1}$, with condition numbers 1, 3, and 10.

The completed checks include direct kinetic and moment references, RK4 replay, preparation and split regeneration, stored-model test re-evaluation, the 225 three-velocity certificates, all saved main-bank scale-map entries and dimensions, four independent direct-quadrature Gramian checks, collision-operator reconstruction, validation-order selection, hidden-direction reidentification, deletion-score replay, 12 multichannel witnesses, and coordinate-covariant symmetry checks. They do not establish independent optimizer convergence, a global minimum state dimension, nonlinear gate performance, or material-specific phonon transport. Two raw HDF5 files, readable model exports, and optimization histories are present locally and identified in the audit manifest. A permanent public repository or archive identifier is not documented in these materials and remains a submission-stage availability item.

\section{General scale-conditioned architecture and prospective extensions}\label{sec:architecture}
This appendix retains the general architectural proposal. The nonlinear gates, shared scale-conditioned encoder, and nonlinear entropy constraints are prospective extensions; the numerical evidence in the main text concerns finite-rank linear models. The following design prescriptions must not be read as a record of additional completed experiments. For a fixed physical model, conditioning parameters are fixed throughout each rollout; the forecast horizon selects rank and fitting objectives, rather than changing a fixed model's generator.

To identify effective transport dynamics without prescribing their macroscopic state variables, we introduce a scale-adaptive latent dynamical network that jointly learns microscopic coarse-graining, state elimination, and reduced evolution. The architecture is constructed so that spatial and temporal scales enter explicitly, whereas the physical interpretation and number of latent variables emerge from the training data. Unlike conventional operator-learning approaches that approximate a predetermined continuum equation, the present formulation seeks the smallest predictive representation consistent with microscopic kinetic evolution.

\subsection{Scale-conditioned latent representation}

Consider a collection of microscopic trajectories
\begin{equation}
\mathcal D
=
\left\{
f^{(n)}(\mathbf x,\lambda,t;\boldsymbol\mu_n)
\right\}_{n=1}^{N_s},
\end{equation}
generated from the BTE under different initial conditions, boundary excitations, and transport regimes. Here \(\boldsymbol\mu_n\) contains the relevant material and scale parameters.

A scale-dependent coarse-graining operator first maps the microscopic distribution onto a prescribed spatial resolution,
\begin{equation}
\bar f_\ell
=
\mathcal C_\ell[f],
\end{equation}
where \(\mathcal C_\ell\) may be implemented through normalized spatial averaging or a conservative multiresolution discretization.

The coarse-grained kinetic state is subsequently encoded into an unconstrained latent representation,
\begin{equation}
	\mathbf z_\theta(\mathbf x,t;\ell)
	=
	\mathcal E_\theta
	\left[
	\bar f_\ell(\mathbf x,\cdot,t),
	\boldsymbol\chi_\ell
	\right],
\end{equation}
where
\begin{equation}
\boldsymbol\chi_\ell
=
(\mathrm{Kn},\mathrm{De},\ell,H)
\end{equation}
denotes the scale-conditioning variables.
The encoder output is
\begin{equation}
\mathbf z_\theta\in\mathbb R^M,
\end{equation}
with \(M\) chosen sufficiently large to represent the relevant kinetic dynamics. No individual latent component is assigned a predetermined interpretation as temperature, heat flux, or a higher-order moment.

The network is therefore required to discover a predictive coordinate system rather than reproduce a manually specified hierarchy of continuum variables.

\subsection{Adaptive state selection}

To determine which latent degrees of freedom remain necessary at each scale, we introduce a scale-conditioned gating operator,
\begin{equation}
	\widetilde{\mathbf z}_\ell
	=
	\mathbf g_\phi(\boldsymbol\chi_\ell)
	\odot
	\mathbf z_\theta,
\end{equation}
where
\begin{equation}
\mathbf g_\phi
=
(g_1,\ldots,g_M),
\qquad
0\leq g_i\leq1.
\end{equation}

Each gate controls the contribution of a latent coordinate to the effective evolution.
For differentiable training, the gates may be represented by continuous stochastic relaxations of binary variables. Their activation probabilities are optimized jointly with the encoder and dynamical model.
The expected active-state dimension is
\begin{equation}
	m_{\mathrm{eff}}(\boldsymbol\chi_\ell)
	=
	\sum_{i=1}^{M}
	\mathbb E[g_i^{\mathrm{binary}}
	\mid\boldsymbol\chi_\ell].
\end{equation}

A sparsity penalty encourages the removal of redundant states,
\begin{equation}
\mathcal L_{\mathrm{dim}}
=
\mathbb E_{\boldsymbol\chi_\ell}
\left[
m_{\mathrm{eff}}(\boldsymbol\chi_\ell)
\right].
\end{equation}

Because latent coordinates are not uniquely identifiable under invertible transformations, individual gate activations are not interpreted directly as unique microscopic modes. The operational dimension is meaningful only at fixed model class, spatial stencil, regularity, precision, observation protocol, and error norm. Use bounded-sensitivity encoders and finite precision to exclude arbitrary information packing. Compare predictive subspaces and their responses, rather than individual latent coordinates.

To prevent information leakage through inactive channels, all downstream evolution and prediction operators receive only the gated state. The encoder and gate are additionally evaluated under capacity-matched architectures and independent initial conditions.

\subsection{Learned effective dynamics}

The retained latent variables evolve according to a scale-dependent dynamical operator,
\begin{equation}
	\partial_t\widetilde{\mathbf z}_\ell
	=
	\mathcal F_\psi
	\left(
	\widetilde{\mathbf z}_\ell,
	\nabla\widetilde{\mathbf z}_\ell,
	\nabla^2\widetilde{\mathbf z}_\ell;
	\boldsymbol\chi_\ell
	\right).
\end{equation}

The operator \(\mathcal F_\psi\) is not constrained to Fourier or CV transport. Instead, it is trained to reproduce the projected evolution generated by the microscopic BTE.
A practical realization separates conservative transport and relaxation,
\begin{equation}
	\partial_t\widetilde{\mathbf z}_\ell
	+
	\nabla\cdot
	\mathbf F_\psi
	(\widetilde{\mathbf z}_\ell;\boldsymbol\chi_\ell)
	=
	\mathbf R_\psi
	(\widetilde{\mathbf z}_\ell;\boldsymbol\chi_\ell).
\end{equation}

The flux network \(\mathbf F_\psi\) represents spatial transport, whereas \(\mathbf R_\psi\) represents unresolved relaxation processes.
This separation provides an inductive bias consistent with kinetic transport while leaving the effective constitutive structure unknown.
For regimes in which instantaneous closure is insufficient, the dynamical model is extended to include memory,
\begin{equation} \label{dynamical memory model}
	\begin{aligned}
		\partial_t\widetilde{\mathbf z}_\ell(t)
		={}&
		\mathcal F_\psi[
		\widetilde{\mathbf z}_\ell(t)]\\
		&+
		\int_0^t
		\mathcal K_\psi(t-s;\boldsymbol\chi_\ell)
		\widetilde{\mathbf z}_\ell(s)\,ds.
	\end{aligned}
\end{equation}

The memory kernel can be represented by a finite set of auxiliary relaxation variables or by a causal neural convolution. Equation~\eqref{dynamical memory model} additionally requires an unresolved-initial-state term analogous to Eq.~\eqref{initial-state contribution}, unless the preparation fixes it to zero or an initialized auxiliary state represents it. Each auxiliary variable and each retained history sample is part of the state-memory budget. A memory model with arbitrarily initialized hidden history is not comparable to a snapshot-initialized Markov model. This extension permits the architecture to distinguish between increasing the number of instantaneous state variables and retaining nonlocal temporal information.

\subsection{Physics-constrained learning}

The network is trained through a combination of microscopic prediction, conservation, coarse-graining consistency, and state-complexity objectives.
The primary multistep prediction loss is
\begin{equation}
\mathcal L_{\mathrm{pred}}
=
\mathbb E_{\mathcal D}
\left[
\sum_{j=1}^{N_t}
w_j
\left\|
\widehat{\mathcal O}_\psi(t_j)
-
\mathcal O[f(t_j)]
\right\|^2
\right],
\end{equation}
where the observables include coarse-grained energy density, heat flux, and selected kinetic moments.
The predicted observables are obtained through a decoder,
\begin{equation}
\widehat{\mathcal O}_\psi
=
\mathcal D_\omega
(\widetilde{\mathbf z}_\ell).
\end{equation}

Although these observables are used for supervision, their corresponding coordinates are not imposed within the latent state.
Energy conservation is enforced through
\begin{equation}
\mathcal L_{\mathrm{cons}}
=
\left\|
\partial_t\widehat e
+
\nabla\cdot\widehat{\mathbf q}
\right\|^2.
\end{equation}

For a closed system with thermodynamically admissible collision dynamics, a suitable entropy-production constraint may additionally be imposed,
\begin{equation} \label{entropy-production constraint}
\mathcal L_{\mathrm{ent}}
=
\left\|
\min(0,\widehat{\dot S}_{\mathrm{tot}})
\right\|^2,
\end{equation}
where \(\widehat{\dot S}_{\mathrm{tot}}\) denotes the total entropy-production rate, including any relevant boundary entropy fluxes. This rate must come from a specified kinetic entropy or its justified approximation. A freely learned scalar called entropy can collapse to a constant and satisfy Eq.~\eqref{entropy-production constraint} without constraining physics. The concrete near-equilibrium construction in Sec.~\ref{sec:linear_realization} uses a fixed entropy Hessian instead.

For two nested coarse-graining scales \(\ell_1<\ell_2\), consistency is encouraged through
\begin{equation}
	\mathcal C_{\ell_2\leftarrow\ell_1}
	\widehat{\mathcal O}_{\ell_1}(t)
	\approx
	\widehat{\mathcal O}_{\ell_2}(t).
\end{equation}

The corresponding loss is
\begin{equation}
\mathcal L_{\mathrm{scale}}
=
\left\|
\mathcal C_{\ell_2\leftarrow\ell_1}
\widehat{\mathcal O}_{\ell_1}
-
\widehat{\mathcal O}_{\ell_2}
\right\|^2.
\end{equation}

This condition links different physical resolutions without requiring their latent coordinates to coincide.
The total objective is
\begin{equation}
	\begin{aligned}
		\mathcal L
		={}&
		\mathcal L_{\mathrm{pred}}
		+
		\lambda_c\mathcal L_{\mathrm{cons}}
		+
		\lambda_s\mathcal L_{\mathrm{scale}}\\
		&+
		\lambda_e\mathcal L_{\mathrm{ent}}
		+
		\lambda_d\mathcal L_{\mathrm{dim}}.
	\end{aligned}
\end{equation}

The regularization parameters control the tradeoff between predictive accuracy, physical consistency, and model complexity.

\subsection{Multiscale architecture and implementation}

The proposed network consists of four coupled modules: a kinetic encoder, a scale-conditioned state gate, a latent evolution operator, and an observable decoder.
At each spatial resolution \(\ell_j\), the microscopic distribution is first mapped onto a conservative coarse-grained representation. A shared encoder then produces \(M\) candidate latent variables. Scale-conditioned gates determine the active state subset, and a neural evolution operator advances the retained variables over the requested time horizon.
The multiresolution hierarchy is defined by
\begin{equation}
\ell_{j+1}=b\ell_j,
\qquad b>1.
\end{equation}

Unlike conventional deep networks, each resolution level has an explicitly defined physical length scale. The number of computational layers within an encoder or evolution block is treated as a numerical hyperparameter, not as a physical observable.
The model is trained using trajectories spanning ballistic, quasiballistic, and diffusive regimes. Training batches sample different initial perturbations, sample lengths, relaxation spectra, and observation times. A curriculum may begin with short-horizon prediction and progressively increase the rollout duration to stabilize the learned dynamics.
For each scale, the active-state dimension is selected using held-out prediction errors,
\begin{equation}
	\widehat m^*_{\mathrm{bank}}(\ell,H,\varepsilon)
	=
	\min
	\left\{
	m:
	\mathcal E_{\mathrm{val}}(m,\ell,H)
	\leq\varepsilon
	\right\}.
\end{equation}

This procedure prevents the sparsity penalty alone from determining the inferred physical complexity. Dimension and all hyperparameters are selected on validation trajectories; the test set is used once after selection. Report the resulting quantity as \(\widehat m^*\), a finite-validation model-bank estimate, rather than a proven global minimum. A passing finite sample alone does not certify a population or worst-case error bound.

\subsection{Identification of emergent transport laws}

The final stage is to identify the physical structure represented by the learned states.
For a spatially homogeneous reference state \(\mathbf z_0\), the learned dynamics are linearized as
\begin{equation}
\partial_t\delta\mathbf z
=
\mathcal J_\ell\delta\mathbf z,
\end{equation}
where
\begin{equation} \label{Frechet derivative}
\mathcal J_\ell
=
\left.
\frac{\delta\mathcal F_\psi}
{\delta\mathbf z}
\right|_{\mathbf z_0}.
\end{equation}

In Eq.~\eqref{Frechet derivative}, \(\mathcal J_\ell\) is the Fr\'echet derivative of the full spatial evolution operator; derivatives with respect to local states alone omit transport. For spatially varying perturbations, Fourier transformation gives
\begin{equation}
\partial_t\delta\widetilde{\mathbf z}
=
\mathcal J_\ell(\mathbf k)
\delta\widetilde{\mathbf z}.
\end{equation}

The eigenvalues
\begin{equation}
	\mathcal J_\ell(\mathbf k)\mathbf r_n
	=
	s_n(\mathbf k;\ell)\mathbf r_n
\end{equation}
define the learned dispersion and relaxation spectrum.
A diffusive mode is identified by the long-wavelength behavior
\begin{equation}
s(\mathbf k)
=
-\alpha_{\mathrm{eff}}|\mathbf k|^2
+
o(|\mathbf k|^2).
\end{equation}

A CV-type two-mode closure is supported when the relevant eigenvalues satisfy, over the tested regime,
\begin{equation}
	\tau_{\mathrm{eff}}s^2+s+
	\alpha_{\mathrm{eff}}|\mathbf k|^2
	\approx0.
\end{equation}

More complicated spectral structures indicate additional relaxation channels or the need for a nonlocal effective description. Eigenvalue agreement alone does not identify a constitutive law: compare energy/flux eigenvectors, observable residues, and responses to independent preparations. A hidden mode with zero observable residue cannot be inferred from that measurement.

To establish a connection with microscopic statistical dynamics, the learned invariant subspaces are compared with the corresponding slow spectral subspaces of the linearized BTE. Their agreement is quantified using projection-operator distances or principal angles rather than componentwise comparisons of latent variables.
The resulting scale-dependent state dimension, dispersion spectrum, and closure error jointly characterize the emergence of effective transport laws.
The working hypothesis is that, for a fixed material and a compatible sequence of observation protocols, kinetic-to-continuum transport admits decreasing predictive complexity. This is neither a universal monotonicity theorem nor a guaranteed BTE--CV--Fourier sequence.

\begin{equation}
\begin{gathered}
 \text{Kinetic dynamics}\\
 \downarrow\\
 \text{Scale-dependent predictive states}\\
 \downarrow\\
 \text{Validated effective transport}
\end{gathered}
\end{equation}
with the retained state space determined by microscopic relaxation, observation scale, and predictive accuracy rather than by a prescribed constitutive equation.

\section{Extended training and validation design}\label{app:protocol}
This appendix preserves the original staged protocol as design guidance. Imperative statements, candidate optimizer settings, requested ablations, and comparisons not explicitly documented in the numerical-results section are proposals, not additional reported results. The completed linear-benchmark settings and the scope of their verification are recorded in Appendix~\ref{app:reporting}; the remaining design prescriptions below are prospective.

\subsection{A staged experiment}
\paragraph{Data and reference solver.}
Begin with Eqs.~\eqref{eq:three}--\eqref{eq:hidden-pair} on a periodic interval using a Fourier solver and exact matrix exponentials for each mode. Use \(N_x=128\) initially and repeat selected cases at 256 points. Set \(c=\tau=1\) as reference units and vary the domain or perturbation wave number so that \(kc\tau\) spans \(10^{-2}\) to \(10\). Generate about 2,000 independently drawn low-amplitude preparations, mixing equilibrium-like perturbations, independent flux perturbations, hidden pairs, and multimode superpositions. Rescale each perturbation to keep every \(f_i\) positive. Initial amplitudes below one percent of the reference population provide a useful first linear-response check, not a universal threshold.

Save densely sampled early transients (for example \(\Delta t=0.01\tau\)) and logarithmically spaced later observations. Diffusive cases require horizons comparable to \((\alpha k^2)^{-1}\), not merely a few collision times. Compute those observations directly with the exact propagator. Test \(t_w/\tau=0,1,5\) separately, with matched observation weights and normalized amplitudes. Conservative cell averaging at several \(\ell/L\) values is applied consistently to both input and output. Do not divide errors by individual late-time signals that decay to zero.

\paragraph{Split and fit.}
Split by complete preparation into 70\% training, 15\% validation, and 15\% test; both members of a pair belong to the same split. Reserve wave-number intervals and, in later multirate models, entire collision spectra for out-of-distribution tests. Fit a linear projection first, then the scale-conditioned network, using multistep energy and flux predictions. \rev{The transparent three-velocity fits shown here use L-BFGS-B with matrix-exponential Fr\'echet derivatives and five optimization seeds. The multichannel comparison uses three learned-model seeds per rank. Adam at learning rate $10^{-3}$, batches of 32 trajectories, and validation-based early stopping are possible choices for the proposed nonlinear extension, rather than the optimizer used to produce these figures.}

For paired preparations, a normalized response loss is
\begin{equation}\label{eq:pairloss}
 \mathcal L_{\mathrm{pair}}
 =\mathbb E\|[\widehat Y_+-\widehat Y_-]-[Y_+-Y_-]\|_W^2.
\end{equation}
This supervises the physical response to unresolved perturbations without assigning a latent coordinate to \(P\). Merely separating every pair in latent space would also preserve irrelevant information and is not the objective. Along with the exact conservation and dissipation construction, use
\begin{equation}
 \mathcal L_{\mathrm{fit}}=\mathcal L_{\mathrm{pred}}
 +\lambda_p\mathcal L_{\mathrm{pair}}
 +\lambda_s\mathcal L_{\mathrm{scale}}.
\end{equation}
Select rank by validation error, including the pair responses, at predeclared tolerances such as \(10^{-2}\) and \(10^{-3}\). Keep the pair weight and normalization fixed across competing models. Require a latent rollout to agree with re-encoding the kinetic state at multiple times; otherwise a good future decoder may still fail as an autonomous state representation.

\paragraph{Active refinement.}
After an initial fit, search for unresolved directions that maximize future output separation while holding the candidate encoded state fixed within a stated tolerance. For the linear model this uses \(\ker U^TG\) and the operator in Eq.~\eqref{eq:obs}; for nonlinear models use constrained perturbation optimization with positivity and fixed thermodynamic norm. Simulate the resulting pairs, add them only to the training set, and refit. This targeted search tests information discarded by the \emph{learned} state, beyond the initial energy-and-flux constraint. Independent test preparations remain untouched.

\paragraph{Extension beyond the toy problem.}
Increase to 16--64 velocity or phonon bins with positive quadrature weights. Use a detailed-balance collision matrix with energy conservation and at least two nonzero relaxation scales before attempting a full material-specific BTE. A naive mode-dependent relaxation-time approximation can violate energy conservation unless its equilibrium temperature is solved consistently. Vary the fast-to-slow relaxation ratio and the observable coupling to each channel independently. The objective is to learn hidden combinations not supplied as conventional moments, and to test their effect on response functions and held-out preparations. A neural model must improve on a transparent linear reduction in this setting before a claim of nonlinear discovery is justified.

\subsection{Controls and measurable outcomes}
At matched field dimension, spatial resolution, and initialization information, compare Fourier and CV closures; fixed moment truncations; PCA with fitted dynamics; finite-horizon observability reduction; a time-lagged autoencoder; the learned projection; and a memory-augmented model with its storage cost reported. Remove pair supervision, entropy structure, and scale conditioning in separate ablations. Reconstruction accuracy alone is not a success criterion.

\begin{revision}
The tests in Sec.~\ref{sec:numerical_results} implement three complementary checks: independent hidden-pair responses, deletion followed by refitting, and comparison of achieved scale-dependent errors with information bounds. A coordinate deletion without refitting would test fragility alone. Here, matched-input certificates identify residual failures due to missing information, while the state-consistency audit prevents a good output decoder from being mistaken for a closed autonomous state. Changes of latent basis across seeds are admissible when the lifted subspaces and responses agree.
\end{revision}

\begin{revision}
The evidence is carried jointly by the paired trajectories, state-removal tests, observable singular spectra, recovered physical profiles, and explicit omitted-memory terms. Seed variation quantifies optimization sensitivity; deterministic parameter scans and information certificates are not independent experimental replicates. The corresponding conclusions remain conditional on the tested preparation ensembles, error norms, and finite model banks.
\end{revision}

\makeatletter
\let\auto@bib@innerbib\@empty
\makeatother


\begin{thebibliography}{99}
\bibitem{Lebowitz1999}
J. L. Lebowitz, Statistical mechanics: A selective review of two central issues,
Rev. Mod. Phys. \textbf{71}, S346 (1999).

\bibitem{Barnett2023}
L. Barnett and A. K. Seth, Dynamical independence: Discovering emergent macroscopic processes in complex dynamical systems,
Phys. Rev. E \textbf{108}, 014304 (2023).

\bibitem{Lebowitz1988}
J. L. Lebowitz, E. Presutti, and H. Spohn,
Microscopic models of hydrodynamic behavior,
J. Stat. Phys. \textbf{51}, 841--862 (1988).

\bibitem{Bertin2006}
E. Bertin, M. Droz, and G. Grégoire,
Boltzmann and hydrodynamic description for self-propelled particles,
Phys. Rev. E \textbf{74}, 022101 (2006).

\bibitem{Ginzburg2007}
I. Ginzburg,
Lattice Boltzmann modeling with discontinuous collision components: Hydrodynamic and advection-diffusion equations,
J. Stat. Phys. \textbf{126}, 157--206 (2007).

\bibitem{Thuroff2014}
F. Thüroff, C. A. Weber, and E. Frey,
Numerical treatment of the Boltzmann equation for self-propelled particle systems,
Phys. Rev. X \textbf{4}, 041030 (2014).

\bibitem{ChapmanCowling1970} S. Chapman and T. G. Cowling, \emph{The Mathematical Theory of Non-Uniform Gases}, 3rd ed. (Cambridge University Press, 1970).

\bibitem{Grad1949}
H. Grad, On the kinetic theory of rarefied gases,
Commun. Pure Appl. Math. \textbf{2}, 331--407 (1949).

\bibitem{Zhu2026JAP}
P. Zhu, H. Zhang, C. Ibarra-Castanedo, X. Maldague, and A. Mandelis,
Making neural networks understand internal heat transfer using Fourier-transformed thermal diffusion wave fields,
J. Appl. Phys. \textbf{140}, 105101 (2026).

\bibitem{Zhu2026NonHermitian}
P. Zhu,
Non-Hermitian exceptional dynamics in first-order heat transport,
arXiv:2604.13639 (2026).

\bibitem{Zhu2026Snell}
P. Zhu, J. Lecompagnon, P. D. Hirsch, M. Ziegler, and A. Mandelis,
The spectral domain Snell law in diffusion-wave fields,
arXiv:2603.24094 (2026).

\bibitem{Chen2021}
G. Chen,
Non-Fourier phonon heat conduction at the microscale and nanoscale,
Nat. Rev. Phys. \textbf{3}, 555--569 (2021).

\bibitem{Guyer1966}
R. A. Guyer and J. A. Krumhansl,
Thermal conductivity, second sound, and phonon hydrodynamic phenomena in nonmetallic crystals,
Phys. Rev. \textbf{148}, 778 (1966).

\bibitem{Cattaneo1958} C. Cattaneo, Sur une forme de l'\'equation de la chaleur \'eliminant le paradoxe d'une propagation instantan\'ee, C. R. Acad. Sci. \textbf{247}, 431--433 (1958).

\bibitem{Kabernik2018}
O. Kabernik,
Quantum coarse graining, symmetries, and reducibility of dynamics,
Phys. Rev. A \textbf{97}, 052130 (2018).

\bibitem{Rahav2007}
S. Rahav and C. Jarzynski,
Fluctuation relations and coarse-graining,
J. Stat. Mech. (2007) P09012.

\bibitem{Feret2009}
J. Feret, V. Danos, J. Krivine, and W. Fontana,
Internal coarse-graining of molecular systems,
Proc. Natl. Acad. Sci. U.S.A. \textbf{106}, 6453--6458 (2009).

\bibitem{Mori1965} H. Mori, Transport, collective motion, and Brownian motion, Prog. Theor. Phys. \textbf{33}, 423--455 (1965).

\bibitem{Zwanzig1961} R. Zwanzig, Memory effects in irreversible thermodynamics, Phys. Rev. \textbf{124}, 983--992 (1961).

\bibitem{Noid2013} W. G. Noid, Perspective: Coarse-grained models for biomolecular systems, J. Chem. Phys. \textbf{139}, 090901 (2013).

\bibitem{Rupe2022}
A. Rupe, V. V. Vesselinov, and J. P. Crutchfield,
Nonequilibrium statistical mechanics and optimal prediction of partially-observed complex systems,
New J. Phys. \textbf{24}, 103033 (2022).

\bibitem{LozanoDuran2022}
A. Lozano-Durán and G. Arranz,
Information-theoretic formulation of dynamical systems: Causality, modeling, and control,
Phys. Rev. Research \textbf{4}, 023195 (2022).

\bibitem{Gross2006}
T. Gross and U. Feudel,
Generalized models as a universal approach to the analysis of nonlinear dynamical systems,
Phys. Rev. E \textbf{73}, 016205 (2006).

\bibitem{Presse2013}
S. Pressé, K. Ghosh, J. Lee, and K. A. Dill,
Principles of maximum entropy and maximum caliber in statistical physics,
Rev. Mod. Phys. \textbf{85}, 1115 (2013).

\bibitem{Asllani2018}
M. Asllani, R. Lambiotte, and T. Carletti,
Structure and dynamical behavior of non-normal networks,
Sci. Adv. \textbf{4}, eaau9403 (2018).

\bibitem{Tishby1999} N. Tishby, F. C. Pereira, and W. Bialek, The information bottleneck method, in \emph{Proceedings of the 37th Allerton Conference on Communication, Control, and Computing}, pp. 368--377 (1999).

\bibitem{Wang2021}
D. Wang and P. Tiwary, State predictive information bottleneck,
J. Chem. Phys. \textbf{154}, 134111 (2021).

\bibitem{Brunton2016} S. L. Brunton, J. L. Proctor, and J. N. Kutz, Discovering governing equations from data by sparse identification of nonlinear dynamical systems, Proc. Natl. Acad. Sci. USA \textbf{113}, 3932--3937 (2016).

\bibitem{Champion2019} K. Champion, B. Lusch, J. N. Kutz, and S. L. Brunton, Data-driven discovery of coordinates and governing equations, Proc. Natl. Acad. Sci. USA \textbf{116}, 22445--22451 (2019).

\bibitem{Wehmeyer2018}
C. Wehmeyer and F. No\'e,
Time-lagged autoencoders: Deep learning of slow collective variables for molecular kinetics,
J. Chem. Phys. \textbf{148}, 241703 (2018).

\bibitem{Mardt2018} A. Mardt, L. Pasquali, H. Wu, and F. No\'e, VAMPnets for deep learning of molecular kinetics, Nat. Commun. \textbf{9}, 5 (2018).

\bibitem{Han2019}
J. Han, C. Ma, Z. Ma, and W. E,
Uniformly accurate machine learning-based hydrodynamic models for kinetic equations,
Proc. Natl. Acad. Sci. U.S.A. \textbf{116}, 21983--21991 (2019).

\bibitem{Schotthofer2022}
S. Schotth\"ofer, T. Xiao, M. Frank, and C. Hauck,
Structure preserving neural networks: A case study in the entropy closure of the Boltzmann equation,
Proc. Mach. Learn. Res. \textbf{162}, 19406--19433 (2022).

\bibitem{Lin2021}
Y. T. Lin, Y. Tian, D. Livescu, and M. Anghel,
Data-driven learning for the Mori--Zwanzig formalism: A generalization of the Koopman learning framework,
SIAM J. Appl. Dyn. Syst. \textbf{20}, 2558--2601 (2021).

\bibitem{Lu2026}
W. Lu, H. Ma, W. Lin, and Q. Zhu,
Interpretable functional Koopman learning with non-Markovian closure for spatiotemporal systems,
Proc. Mach. Learn. Res. \textbf{306}, 81278--81322 (2026).
\end{thebibliography}
\end{document}